\documentclass[11pt]{article}
\usepackage{a4}

\usepackage{amsfonts}
\usepackage{amsmath}
\usepackage{amssymb}
\usepackage{graphicx}
\usepackage{cite}
\usepackage{cancel}
\newcommand{\tH}{{\widetilde{H}}}
\newcommand{\hY}{{\hat{Y}}}
\newcommand{\hx}{{\hat{\xi}}}
\newcommand{\he}{{\hat{\eta}}}
\newcommand{\hps}{{\hat{\psi}}}
\newcommand{\hC}{{\hat{C}}}
\newcommand{\hph}{{\hat{\phi}}}
\newcommand{\bea}{\begin{eqnarray}}
\newcommand{\eea}{\end{eqnarray}}
\newcommand{\be}{\begin{equation}}
\newcommand{\ee}{\end{equation}}
\newcommand{\diag}{{\rm diag}}
\newcommand{\del}{\partial}
\newcommand{\nn}{\nonumber}

\newcommand{\Tr}{{\rm Tr}}

\newcommand{\e}{\epsilon}

\newcommand{\tr}{{\rm tr}}

\newcommand{\Z}{\mathbb{Z}}
\newcommand{\R}{\mathbb{R}}

\makeatletter

\@addtoreset{equation}{section}
\makeatother

\def\href#1#2{#2}

\newcommand{\beqs}{\begin{equation*}}  
\def\beq{\begin{equation}}

\newcommand{\eeqs}{\end{equation*}}
\def\eeq{\end{equation}}

\newcommand{\beqas}{\begin{eqnarray*}}
\newcommand{\beqa}{\begin{eqnarray}}

\newcommand{\eeqas}{\end{eqnarray*}}
\newcommand{\eeqa}{\end{eqnarray}}     

\begin{document}
\begin{titlepage}

\setcounter{page}{0}
\renewcommand{\thefootnote}{\fnsymbol{footnote}}

\begin{flushright}
TIFR/TH/12-16\\
\end{flushright} 
\vspace{10mm}

\begin{center}
{\large\bf
An anisotropic hybrid non-perturbative formulation for 
4D ${\cal N} = 2$ supersymmetric Yang-Mills theories.}

\vspace{10mm}
Tomohisa Takimi%
\footnote{\tt takimi@theory.tifr.res.in}  
\vspace{10mm}

{\em 
Department of Theoretical Physics, Tata Institute of Fundamental Research, Homi Bhabha Road, Mumbai
400 005, India 
}
\end{center}

\vspace{20mm}
\centerline{{\bf Abstract}}
\vspace{3mm}
We provide a simple non-perturbative formulation 
for non-commutative 
four-dimensional ${\cal N} = 2$ supersymmetric Yang-Mills theories.
The formulation is constructed by a combination of
deconstruction (orbifold projection), 
momentum cut-off and matrix model techniques.
We also propose a moduli fixing term that preserves lattice supersymmetry on
the deconstruction formulation.
Although the analogous formulation for four-dimensional ${\cal N} = 2$ 
supersymmetric Yang-Mills theories is proposed also in \cite{Hanada:2011qx},
our action is simpler and 
better suited for computer simulations.
Moreover, not only for the non-commutative theories, 
our formulation has a potential to be a non-perturbative tool
also for the commutative four-dimensional ${\cal N} = 2$ supersymmetric 
Yang-Mills theories.
\end{titlepage}

\newpage
 \tableofcontents
\setcounter{footnote}{0}
\section{Introduction}
It is very important to find promising 
numerical frameworks for non-perturbative studies of 
supersymmetric gauge theories.
Although lattice formulations are natural candidates,
it is not so straightforward to apply them for the purpose.
This is because there tends to be many parameters which must be fine-tuned
due to the SUSY breaking by the lattice cut-off. 

So far, ${\cal N} =1$ pure supersymmetric Yang-Mills theories 
{\it without} scalar fields in three 
\cite{Maru:1997kh} and four \cite{Giedt:2009yd} dimensions have been 
shown to be
free from fine-tunings.
For one-dimensional supersymmetric matrix quantum mechanics, a "non-lattice"
technique is applicable \cite{Hanada:2007ti}.
By using this method, supersymmetric matrix quantum mechanics has been 
extensively investigated. In particular, quantitative agreement
with the gauge/gravity duality conjecture has been obtained \cite{Anagnostopoulos:2007fw,Hanada:2008gy,Hanada:2008ez,Hanada:2009ne,Hanada:2011fq}
\footnote{There are also other 
numerical studies \cite{Berenstein:2008jn,Berenstein:2010xw} 
in the context of gauge/gravity duality.},
and qualitative consistency with the lattice calculations
are also obtained in \cite{Catterall:2008yz,Catterall:2009xn}.
%
%

However, 
in supersymmetric field theories {\it with} scalar fields, there tends to be a
plethora of the relevant operators violating supersymmetry
whose coefficients must be fine-tuned.
For example, 
scalar mass terms are difficult to
exclude without fermionic symmetries. 
To overcome such difficulties, several lattice formulations
which preserve a partial SUSY on the lattice, are proposed in~\cite{Catterall:2009it,Cohen:2003xe,Cohen:2003qw,Kaplan:2005ta,Sugino:2003yb,Sugino:2004qd,Sugino:2004uv,Sugino:2006uf,Catterall:2004np,DAdda:2005zk,Joseph:2011xy}.
Although 
these models 
have succeeded to be free from fine-tunings in two or three-dimensional cases
thanks to the super-renormalizability,
extended supersymmetric four-dimensional theories with a finite rank of 
gauge group were out of 
reach.\footnote{
The first attempt to construct a lattice formulation for the ${\cal N} =2$
theory is discussed in~\cite{Elitzur:1982vh}.
An ${\cal N} = (2,2)$ $SU(N)$ formulation of the orbifold lattice gauge theory is discussed in \cite{Kanamori:2012et}. There is also another approach 
without employing the exact SUSY on the lattice \cite{Suzuki:2005dx}.
And there are several 
numerical studies on the two-dimensional 
theories~\cite{Kanamori:2008bk,Hanada:2009hq,Hanada:2010qg,Catterall:2010fx}. 
}
\footnote{For the large $N$, in the planar limit, four-dimensional theories 
can be obtained 
\cite{Ishii:2008ib,Ishiki:2011ct} by using the large $N$ reduction
\cite{Eguchi:1982nm}.}
\footnote{In \cite{Catterall:2009it}, 
the number of fine-tunings in an ${\cal N} =4$ four-dimensional 
lattice model has been estimated.
And the number of fine-tunings has turned out to be 1 by one-loop
perturbative calculations in \cite{Catterall:2011pd}.}

The method we will use to avoid fine-tunings in four dimensions is
to introduce {\it anisotropic regularization}.
This approach was used by 
Hanada et al.~\cite{Hanada:2010kt,Hanada:2010gs,Hanada:2011qx}, 
who first obtained non-perturbative four-dimensional 
formulations free from fine-tunings.
They considered the lattice regularization of
the mass-deformed two-dimensional supersymmetric
Yang-Mill theory, which provides an additional two dimensions 
as a Fuzzy 
2-sphere~\cite{Myers:1999ps,Maldacena:2002rb}.\footnote{The 
construction of the four-dimensional non-commutative
space from the zero 
dimensions has been discussed in \cite{Unsal:2005us,Ydri:2007ua},
} 
They performed lattice regularization 
along the original two dimensions 
by utilizing the balanced topological field theory formalism 
in \cite{Sugino:2003yb,Dijkgraaf:1996tz}. 
On the lattice, a partial SUSY is preserved.
The emergent Fuzzy 2-sphere directions are regularized 
by the non-commutative parameter $\Theta$ 
(and thus UV cut-off $\hat{\Lambda}$) and 
the Fuzzy sphere radius $R_f$.
To take the target four-dimensional continuum limit with no fine-tunings,
they take the following steps:
\begin{enumerate}
\item Taking the continuum limit along the original two-dimensional lattice 
directions. 
\item After that, 
taking the decompactified limit of
the Fuzzy sphere,
$R_f \to \infty (\hat{\Lambda} \to \infty)$.
\item Finally taking commutative limit $\Theta \to 0$.
\end{enumerate}
During the first step, 
the theory is regarded as a two-dimensional theory
with super-renormalizability.
Hence the target intermediate continuum theory can be reached 
without fine-tunings, where two of the four dimensions of the 
intermediate theory are kept as a Fuzzy 2-sphere with finite 
$\Theta, \hat{\Lambda}, R_f$.
And during the subsequent steps, 
we can use symmetries 
restored by the first step 
to circumvent dangerous corrections.
By this series approach, 
we can finally get the target continuum limit without fine-tunings.

From this study, 
we can see following advantages of anisotropic formulations 
with taking the limits in a stepwise manner:\footnote{Actually, 
the potency of the anisotropic continuum limit has been 
already mentioned in \cite{Kaplan:2002wv} 
to reduce the number of fine-tunings.}
\begin{itemize}
\item In the first step, the theory can be regarded as a low-dimensional
theory. 
Thanks to the super-renormalizability,
it is easy to get a desirable intermediate theory 
from which 
we can approach the final target theory.
\item In later steps, symmetries recovered in earlier steps help
to protect from dangerous corrections.
\end{itemize}

In this paper, 
we will construct a
non-perturbative formulation for 
the non-commutative 
four-dimensional ${\cal N} = 2$ supersymmetric Yang-Mill theories
by further developing such an anisotropic treatment.
Although an analogous formulation is constructed based on the BTFT formalism
in \cite{Hanada:2011qx} in a beautiful way,
the formulation is rather complicated and not so easy to put on a
computer.
To construct a more convenient action, 
we employ a combination of orbifolding~\cite{Cohen:2003xe,Cohen:2003qw,Kaplan:2005ta}, momentum cut-off~\cite{Hanada:2007ti,Anagnostopoulos:2007fw,Hanada:2008gy,Hanada:2008ez,Hanada:2009ne,Hanada:2011fq,Hanada:2010jr} and the
Fuzzy sphere techniques. 
We start from the mass-deformed 
one-dimensional supersymmetric matrix model
with 8 supercharges \cite{Kim:2006wg}, 
and we perform the orbifold projection on it
with keeping the one dimension continuum.
After that the continuum dimension is regularized 
by the momentum cut-off,
then we obtain the action for the 
mass-deformed two-dimensional supersymmetric 
Yang-Mills theory on 
$\R^1_{\Lambda} \times \R^1_{orb}$, 
where 
$\R^1_{\Lambda}$ 
is the one dimensions regularized by the momentum cut-off
and 
$\R^1_{orb}$ is 
the regularized one dimensions which is generated by the orbifolding.
For later use, 
we will call the two-dimensional theory regularized by the 
hybrid of the orbifolding and the momentum cut-off as
the "hybrid regularization theory".
The construction of our formulation is completed by
uplifting the hybrid regularization theory 
to the four dimensions by taking the Fuzzy sphere solution, 
which is employed in the same way as \cite{Hanada:2011qx}.
Basically our formulation  
is obtained by replacing the lattice regularization in \cite{Hanada:2011qx}
with the hybrid regularization of orbifolding and momentum cut-off.
Thus our construction uses 
$\R^1_{\Lambda} \times \R^1_{orb} \times \text{Fuzzy}\,\, S^2$,
as compared to \cite{Hanada:2011qx}'s 
$\R^2_{lat}  \times \text{Fuzzy}\,\, S^2$.
Clearly our theory is more anisotropic than \cite{Hanada:2011qx}.
Remarkable properties and advantages of our approach are:
\begin{itemize}
\item 
There has not been an ${\cal N} =2$
four-dimensional supersymmetric lattice gauge theory
based on the orbifold projection 
because there are too few scalar fields.
In our formulation, 
the large number of scalar
fields is not required, 
since the orbifold projection is performed
along only one direction.
Thus we are able to get ${\cal N} =2$ four-dimensional theories
for the first time.\footnote{
A lattice formulation of four-dimensional N=2 supersymmetric gauge theory
is constructed by Sugino~\cite{Sugino:2003yb}. 
Essentially the same formulation is obtained
by Damgaard and Matsuura~\cite{Damgaard:2007xi} 
by combining orbifolding and the truncation
technique developed in~\cite{Takimi:2007nn}.
}
\item Since the orbifold projection is performed along only one direction,
$8 \times \left( \frac{1}{2}\right)^1 =4$ supercharges can be preserved.
The theory has 
more supersymmetry than \cite{Hanada:2011qx}.
\item 
We can introduce a moduli stabilizing mass term 
that preserves 2 of 4 preserved SUSY on the lattice.
This is the first moduli fixing term without breaking lattice SUSY.
\item 
Because our formulation is constructed by the 
orbifold projection from the one-dimensional matrix model,
it can be easily embedded 
into the one-dimensional matrix model with 8 supercharges.
A matrix model with 8 supercharges 
is already numerically studied in \cite{Hanada:2010jr}.
So it would be easy to apply their numerical 
technique to this formulation.
\end{itemize}

The outline of this paper and the basic steps to 
reach the target theory are described in the next two subsections. 

\subsection{Basic steps 
to reach the target theory and usage of our formulation}
\label{Sec:steps}
This section aims to 
clarify the order of taking limits 
to reach the target theory 
without fine-tunings.
We start with a mass-deformed $U(klN)$ one-dimensional matrix model, 
with mass-deformed parameter $M$ and 8 supercharges,
which we refer to as the mother theory.
The four dimensions in the formulation are 
regularized different manners, namely:
\begin{itemize}
\item $\R^1_{\Lambda}$: Regularized by the momentum cut-off $\Lambda$. 
\item $\R^1_{orb}$: Regularized by orbifold projection with lattice spacing 
$a$, and its number of site is characterized by $N$. 
The moduli stabilizing mass terms are also introduced with 
parameter $\nu_1$.
\item Fuzzy $S^2$: Regularized by UV cut-off $\hat{\Lambda} \sim \frac{Ml}{3}$
, radius $R_f \sim 3/M$, and the non-commutative parameter $\Theta \sim 1/(M^2 l)$.
\end{itemize}
Among these, 
$\R^1_{\Lambda} \times \R^1_{orb}$ corresponds to 
$\R^2_{lat}$ of $\R^2_{lat} \times \text{Fuzzy}\,\, S^2$ 
in \cite{Hanada:2011qx}.
In \cite{Hanada:2011qx}, to avoid the fine-tunings,
the continuum limit of the lattice gauge theory on $\R^2_{lat}$ is taken first 
before managing Fuzzy\,\,$S^2$.
This indicates that we need to 
manage $\R^1_{\Lambda} \times \R^1_{orb}$ 
before Fuzzy\,\,$S^2$.
In our more anisotropic formulation, 
we have to choose between $\R^1_{\Lambda}$ or $\R^1_{orb}$ as the first
direction to be managed. 
When choosing the order in which to relax regulators, we should 
start with a crude regularization which breaks much symmetry. 
This is because such a regulator is easiest to manage in 
an early low-dimensional stage, when the theory will be super-renormalizable.
In our set-up, the momentum cut-off $\Lambda$ breaks both supersymmetry
and gauge symmetry while the orbifolding can preserve a partial SUSY 
and gauge symmetry.
Thus we deal with the momentum cut-off first.
Then we will take the following steps
to get the target 
non-commutative ${\cal N} = 2$ four-dimensional 
supersymmetric Yang-Mills continuum limit without any fine-tunings: 
\begin{enumerate}
\item Taking the $\Lambda \to \infty$ keeping other parameters fixed.
\item $a\to 0$, $N\to 0$ with 
keeping $aN$, $\nu_1$ and regularization parameters of Fuzzy sphere fixed.
\item $aN \to \infty$, $\nu_1 \to 0$ with $k, l, (m =kl), M, \Theta, R_f, 
\hat{\Lambda}$ fixed.
\item $l \to \infty$ ($R_f, \hat{\Lambda} \to \infty$, $M\to 0$) 
with $\Theta, k$ fixed.
\end{enumerate}

At the 1st step, 
since the system can be regarded as a one-dimensional 
system without UV divergences, 
both gauge symmetry and a part of supersymmetry are 
automatically
recovered only by taking the $\Lambda \to \infty$.
Without any fine-tunings,
the orbifold lattice gauge theory on 
$\R^1 \times \R^1_{orb} \times \text{Fuzzy}\,\,S^2$
is obtained as an intermediate theory
from our non-perturbative formulation on 
$\R^1_{\Lambda} \times \R^1_{orb} \times \text{Fuzzy}\,\,S^2$.
The orbifold lattice theory has 4 SUSY in the UV region and 
the 2 of 4 supercharges are exactly preserved on the lattice.
We will explain the 1st step in Sec.~\ref{sec:1d correction}.
As the 2nd step, we will take the continuum limit $a \to 0$
with keeping the moduli stabilizing parameter $\nu_1$ and the volume $aN$ 
fixed. 
After that we take the 3rd step $\nu_1 \to 0, aN \to \infty$.
By the 2nd and 3rd steps, without any fine-tunings, 
we will obtain the non-commutative
supersymmetric 
Yang-Mills theory with 8 SUSY on 
$\R^2 \times \text{Fuzzy}\,\, S^2$.
Full 8 supersymmetry and $SO(2)$ rotational symmetry on $\R^2$ 
are recovered in these steps. 
We will explain the 2nd and 3rd steps in Sec.~\ref{Sec:orb-continuum}. 
By the 4th step,
from the theory on 
$\R^2 \times \text{Fuzzy}\,\, S^2$,
we will obtain the final target theory 
which is the non-commutative ${\cal N} =2$ 
four-dimensional supersymmetric Yang-Mills theory
on 
$\R^2 \times \R^2_{\Theta}$.
Here the $\R^2_{\Theta}$ is the two-dimensional 
non-commutative Moyal plane with the non-commutative parameter $\Theta$.
The theory on $\R^2 \times \text{Fuzzy}\,\, S^2$ 
is connected smoothly to the one on
$\R^2 \times \R^2_{\Theta}$.
We will explain the 4th step in Sec.~\ref{Sec:fourth-step}.
In the above way, we will get the target theory 
by these 4 steps without any fine-tunings.
We also illustrate these steps 
by Table~\ref{figure}.
\begin{table}[htbp]
\begin{center}
\scalebox{0.75}
{ 
\begin{tabular}{c|c|c|c|c|c|cc|c|c|c|c}
\cline{2-4}
\cline{9-11}
& \multicolumn{2}{c}{4d space} & &
\multicolumn{2}{c}{} & &
& \multicolumn{2}{c}{Symmetries} &
\\
\cline{2-4}
\cline{6-6}
\cline{9-11}
&$\R^1_{\Lambda}$ & $\R^1_{orb}$ & 
$S^2_{f}$ 
& &
Theories &
 & & 
$Q$ & Gauge & 
$SO(2)$ & 
\\
\cline{2-4}
\cline{6-6}
\cline{9-11}
& & & & &
Our non-perturbative formulation &
& & & & &
\\
&r & r& 
$S^2_{f}$ & & (Hybrid regularization theory  &
& & 0 & $\times$ & $\times$ &
\\
& & & & &  with Fuzzy $S^2$ solution) &
& & & & &
\\
\cline{2-4} \cline{6-6}
\cline{9-11}
\multicolumn{12}{c}{}
\\
\multicolumn{5}{c}{} &
\multicolumn{7}{l}{
{\Huge \quad $\Downarrow$} \quad
\, $(1)\, \Lambda \to \infty$ \quad \qquad \qquad
{\small \,2 SUSY $\&$ Gauge sym. are recovered.}
}
\\
\multicolumn{12}{c}{}
\\
\cline{2-4}
\cline{6-6}
\cline{9-11}
 &c &r &$S^2_{f}$ & &
Orbifold lattice theory on 
$\R^1 \times \R^1_{orb} \times S^2_f$ &
 & & 2 (UV 4)
& $\circ$ & $\times$ & 
\\
\cline{2-4}
\cline{6-6}
\cline{9-11} 
\multicolumn{12}{c}{}
\\
\multicolumn{5}{c}{} &
\multicolumn{7}{l}{
{\Huge \quad $\Downarrow$} \quad
$
\begin{array}{l}
(2)\, a \to 0 \\
(3)\, \nu_1 \to 0, aN \to \infty
\end{array}
$\quad  {\small 8 SUSY $\&$ $SO(2)$ rotational sym. are recovered.}}
\\
\multicolumn{12}{c}{}
\\
\cline{2-4}
\cline{6-6}
\cline{9-11} 
& \multicolumn{1}{c}{$\R^2$}& &$S^2_{f}$ & &
${\cal N} =2$ 4d Non-commutative $U(k)$ SYM on $\R^2 \times S^2_{f}$ & & & 8
& $\circ$ & $\circ$ & 
\\
\cline{2-4}
\cline{6-6} 
\cline{9-11} 
\multicolumn{12}{c}{}
\\
\multicolumn{5}{c}{} &
\multicolumn{7}{l}{
{\Huge \quad $\Downarrow$} \quad
\, $(4)\, l, \to \infty$
($R_f, \hat{\Lambda} \to \infty$, $M\to 0$) 
with $k,\Theta$: fixed}
\\
\multicolumn{12}{c}{}
\\
\cline{2-4}
\cline{6-6}
\cline{9-11} 
& \multicolumn{1}{c}{$\R^2$}& &$\R^2_{\Theta}$ & &
${\cal N} =2$ 4d Non-commutative $U(k)$ SYM on $\R^2 \times \R^2_{\Theta}$ &
 & & 8
& $\circ$ & $\circ$ & 
\\
\cline{2-4}
\cline{6-6} 
\cline{9-11} 
\multicolumn{12}{c}{}
\\
\hline
\multicolumn{12}{l}{$S_f^2:$ Fuzzy $S^2$ \qquad 
r: The direction is regularized \qquad 
c: The direction is continuum}
\\
\multicolumn{12}{l}{$Q$: The number of supercharges \qquad
$SO(2)$: Rotational symmetry on $\R^2$}
\\
\multicolumn{12}{l}{$\circ$: The symmetry is restored \qquad 
$\times$: The symmetry is still broken}
\end{tabular}}
\end{center}
\caption{The flowchart of steps to take the target theory limit.}
\label{figure}
\end{table}



${}$

Our formulation is a powerful non-perturbative tool for the 
${\cal N} =2 $ non-commutative four-dimensional 
supersymmetric Yang-Mills theories.
Non-commutative gauge theory is an important subject of research 
in order to clarify non-perturbative aspects of gauge theories.
For example, the singularity in the instanton moduli space 
is resolved by the non-commutativity.
So numerical studies of non-commutative four-dimensional 
supersymmetric Yang-Mills theories by using our formulation 
will give a
strong instrument to reveal the non-perturbative structure of supersymmetric
gauge theories.

Not only for the non-commutative gauge theories, 
our formulation has a potential
to be a non-perturbative tool also for the commutative 
four-dimensional ${\cal N}=2$ supersymmetric Yang-Mills theories.
Although the non-commutative four-dimensional ${\cal N} =2$ 
supersymmetric 
Yang-Mills theory is expected not to be continuously connected to the
commutative ${\cal N} =2$ theory~\cite{Minwalla:1999px}, 
there is a discussion
that the non-commutative 
${\cal N}=2$ theory 
may flow to the ordinary commutative theory in the infrared
\cite{Armoni:2001br}.
Since still there is a possibility that $\Theta \to 0$ smoothly connects to 
commutative theory in the infrared region, 
the formulation has a potential to be a non-perturbative formulation also
for the commutative gauge theories.

\subsection{Outlines of this paper}
The outline of this paper is as follows:
In section 2 we will explain how to construct
our non-perturbative formulation which is 
the hybrid regularization theory with the Fuzzy 2-sphere solution.
In section 3 we will explain how to take the target
${\cal N} =2$ non-commutative supersymmetric Yang-Mills limit
on $\R^2 \times \R^2_{\Theta}$.
The absence of fine-tunings in the steps is explained in this 
section.
Section 4 is the summary.

\section{How to construct the non-perturbative formulation}
Here we will explain how to construct our formulation.
We start from the mass-deformed one-dimensional matrix model
with 8 supersymmetry.
First we obtain the orbifold lattice theory on $\R^1 (\text{continuum}) \times 
\R^1_{orb}$
by performing the orbifold projection on the matrix model.
In the subsection \ref{sec:orbifold}, we will explain the orbifold lattice gauge theory.
Second we perform the momentum cut-off regularization on the orbifold action,
then we get the hybrid regularization theory on $\R^1_{\Lambda}\times 
\R^1_{orb}$. We explain the momentum cut-off regularization 
in Sec.~\ref{Momentum cut-off}.
And to complete the construction of our formulation, we expand the 
hybrid regularization theory around the Fuzzy 2-sphere solution.
We explain it in Sec~\ref{Uplift}.
\subsection{Two-dimensional orbifold lattice gauge theory with keeping 
one dimension continuum}
\label{sec:orbifold}
\subsubsection{Mother theory}
We start from the following Euclidean one-dimensional $U(mN)$ matrix model
with 8 supersymmetry, which we call as the "mother theory",
\begin{align}
S^{mat} = \frac{2}{g^2}\int dx_1 \, \Tr \biggl[& \frac{1}{2}D_{1}X^ID_{1}X^I
-\frac{1}{4}[X^I, X^J]^2 
+\frac{i}{2}\Psi^T \gamma^K [X_K, \Psi]
+\frac{1}{2}\Psi^T D_1 \Psi
\nn \\
&-i\frac{M}{6}{\Psi}^T\gamma^{23}{\Psi} 
+\frac{iM}{3}{X}^3D_1{X}^2 +\frac12\left(\frac{M}{3}\right)^2 ({X}^a)^2
 +i\frac{M}{3}\epsilon_{abc}{X}^a{X}^b{X}^c
\biggr] \label{Def:1d mother theory}
\end{align}
where $I,J,K = 2,\ldots 6$ and $a = 4,5,6$,
labeling the bosonic fields $X^I, X^a$.
The integer $m$ is defined as $m = kl$.
$M$ is the mass deformation parameter.
Here we compactify the $x_1$ direction with the 
length $R_1$, $x_1 \sim x_1 + R_1$.
And we impose the periodic boundary condition on all fields.
The fermion $\Psi$ is an 8-component spinor
and each component of the $\Psi$ is written by the 
$\Psi^{(0)}$ as
\begin{align} 
\Psi^{(0)}=\left(\psi_{+1},\psi_{+2},\chi_+,\frac{1}{2}\eta_+, 
\psi_{-1},\psi_{-2},\chi_-,\frac{1}{2}\eta_-\right)^T, 
\label{Fermion reps}
\end{align}
with
\begin{align}
 \Psi&=U_8 \Psi^{(0)}, \label{def:fermion components}
\end{align}
and $U_8$ is
\begin{align}
 U_8&=\frac{1}{2}\left(\begin{matrix}
0&0&-1&i&1&i&0&0\\
1&-i&0&0&0&0&-1&i \\
0&0&i&1&i&-1&0&0 \\
-i&-1&0&0&0&0&-i&-1 \\
1&i&0&0&0&0&1&i \\
0&0&1&i&1&-i&0&0 \\
-i&1&0&0&0&0&i&-1 \\
0&0&-i&1&i&1&0&0
\end{matrix}\right). 
\label{Uf}
\end{align}
Here the all fields are expanded by a basis of the representation 
$T^{\tilde{a}}$ ($\tilde{a} = 1, \cdots, \text{dim}(u(mN))$) as
$\Psi = \Psi^{\tilde{a}} T^{\tilde{a}}, \ldots$, and they are in 
the adjoint representation of the $U(mN)$ gauge group. Hence, they are
$mN \times mN$ matrix valued quantities.
Here $g$ is the one-dimensional gauge coupling with mass dimension $3/2$.
The gamma matrices $\gamma_K$ are 
written in the appendix \ref{Sec:gamma matrix}.
The covariant derivative $D_1$ is defined as
$D_{1} = \del_{x_1} + i [v^1, \cdot\,]$ where the $v^1$ is a gauge field along 
the $x_1$ direction. 
The action (\ref{Def:1d mother theory}) is derived from the 8 supersymmetry analogue \cite{Kim:2006wg} of the 
plane wave matrix model \cite{Berenstein:2002jq}.\footnote{PP wave matrix 
strings are discussed in 
\cite{Sugiyama:2002rs,Das:2003yq}.}
It
is also obtained by the dimensional reduction from the 
mass-deformed Euclidean two-dimensional supersymmetric Yang-Mills theory 
with 8 supercharges, which is described in eq.~(A.28) of \cite{Hanada:2011qx}.

The mother theory 
(\ref{Def:1d mother theory}) 
preserves the following supersymmetry with 8 supercharges,
\bea
\delta v^1 & = & \epsilon^T \Psi, \nn \\
\delta X^I  & = & \epsilon^T \gamma_I\Psi, \nn \\
\delta \Psi & = & \left( -(D_1 X^I)\gamma_I 
+\frac{i}{2} [X^I, X^J]\gamma_{IJ} -\frac{M}{3} X^a\gamma_a
 \gamma_{456}\right)  \epsilon. 
\label{dyn_SUSY_2dE} 
\eea
The SUSY parameter $\epsilon$ is independent of the coordinate $x_1$
while the SUSY parameter
in the plane wave matrix model \cite{Berenstein:2002jq}
depends on the coordinate $x_1$.

\subsubsection{Balanced topological field theory form of the mother theory}
Among the 8 supercharges of the mother theory
(\ref{dyn_SUSY_2dE}),
we pick up two supercharges $Q_+$ and $Q_-$. 
Each $Q_{+}$ and $Q_{-}$ is associated with the parameter $\e_{+}$ and
$\e_{-}$ respectively. The $\e_{\pm}$ are
\begin{equation}
 \e_\pm^T = \e_\pm^{\prime T} U_8^{-1}, 
 \label{epsilon12}
\end{equation}
with
\begin{equation}
 \e'_+ = \left(\begin{matrix}
\varepsilon_+ \\ 0 \\ 0 \\ 0 \\ 0 \\ 0 \\ 0 \\ 0
\end{matrix}\right), \qquad
 \e'_- = \left(\begin{matrix}
0 \\ 0 \\ 0 \\ 0 \\ \varepsilon_- \\ 0 \\ 0 \\ 0
\end{matrix}\right),
\quad (\varepsilon_\pm\text{ : Grassmann numbers}).
\label{epsilon12-2} 
\end{equation}
For later use, we define following complexified fields as
\begin{equation}
Y=iX_2+X_3, \qquad Y^\dag= -iX_2+X_3,
\label{def:complex-boson-matrix}
\end{equation} 
\begin{equation}
C = 2X^4, \qquad \phi_{\pm} = X^5\pm i X^6 ,
\end{equation}
\begin{equation}
\xi_{\pm } =  i\psi_{\pm 2} + \chi_{\pm},
\qquad
\xi_{\pm }^{\dag} = - i\psi_{\pm 2} + \chi_{\pm}.
\end{equation}
Off-shell $Q_{\pm}$ transformations are
\begin{align}
 Q_\pm v_1&=\psi_{\pm 1}, \quad
 Q_\pm \psi_{\pm 1}=\pm iD_1 \phi_\pm, \quad
 Q_\mp \psi_{\pm 1}=\frac{i}{2} D_1 C \mp \tH_1, \nn \\
 Q_\pm \tH_1 &=
 [\phi_\pm,\psi_{\mp 1}]\mp \frac{1}{2}[C,\psi_{\pm 1}]
 \mp\frac{i}{2}D_1 \eta_\pm +\frac{M}{3}\psi_{\pm 1},  \nn \\
 Q_\pm X_2&=\psi_{\pm 2}, \quad
 Q_\pm \psi_{\pm 2}=\pm [\phi_\pm, X^2], \quad
 Q_\mp \psi_{\pm 2}=\frac{1}{2} [C,X^2] \mp \tH_2, \nn \\
 Q_\pm \tH_2 &=
 [\phi_\pm,\psi_{\mp 2}]
 \pm\frac{1}{2}[X^2, \eta_\pm] 
\mp \frac{1}{2}[C,\psi_{\pm 2}]
+\frac{M}{3}\psi_{\pm 2},  \nn \\
 Q_\pm X^3 &= \chi_\pm, \quad 
 Q_\pm \chi_\pm = \pm[\phi_\pm,X^3], \quad
 Q_\mp \chi_\pm = \frac{1}{2}[C,X^3] \mp H, \nn \\
 Q_\pm H &= [\phi_\pm,\chi_\mp] \pm\frac{1}{2}[X^3,\eta_\pm]
\mp \frac{1}{2}[C,\chi_\pm] + \frac{M}{3}\chi_\pm,
 \label{Qpm-M1}
\end{align}
\begin{align}
 Q_\pm Y&=\xi_{\pm }, \quad
 Q_\pm \xi_{\pm }=\pm [\phi_\pm, Y], \quad
 Q_\mp \xi_{\pm }=\frac{1}{2} [C, Y] \mp H_y, \nn \\
 Q_\pm H_y &=
 [\phi_\pm,\xi_{\mp }]
 \pm\frac{1}{2}[Y, \eta_\pm] 
\mp \frac{1}{2}[C,\xi_{\pm }]
+\frac{M}{3}\xi_{\pm },  \nn \\
 Q_\pm Y^\dag &= \xi_\pm^\dag, \quad 
 Q_\pm \xi_\pm^\dag = \pm[\phi_\pm, Y^{\dag}], \quad
 Q_\mp \xi_{\pm }^\dag=\frac{1}{2} [C, Y^\dag] \mp H_y^\dag, \nn \\
 Q_\pm H_y^\dag &=
 [\phi_\pm,\xi_{\mp }^\dag]
 \pm\frac{1}{2}[Y^\dag, \eta_\pm] 
\mp \frac{1}{2}[C,\xi_{\pm }^\dag]
+\frac{M}{3}\xi_{\pm }^\dag,  \nn \\
 Q_\pm C &= \eta_\pm, \quad
 Q_\pm \eta_\pm = \pm[\phi_\pm,C]+\frac{2M}{3}\phi_\pm, \quad
 Q_\mp \eta_\pm = \mp[\phi_+,\phi_-]\pm\frac{M}{3}C, \nn \\
 Q_\pm \phi_\pm &= 0, \quad
 Q_\mp \phi_\pm = \mp \eta_\pm.
 \label{Qpm-M2}
\end{align}
Here 
\begin{equation}
H_y = i \tilde{H}_2 + H, \qquad 
H_y^\dag = -i \tilde{H}_2 + H,
\end{equation}
and $\tH_1$ are auxiliary fields.

The mother theory action (\ref{Def:1d mother theory})
is also invariant under the $SU(2)_R$ transformations
whose generators are
\begin{align}
 J_{++}=&\int d x_1 \biggl[
\psi^{\tilde{a}}_{+ 1}(x_1)\frac{\delta}{\delta \psi^{\tilde{a}}_{-1}(x_1)}
+\xi^{\tilde{a}}_{+}(x_1)\frac{\delta}{\delta \xi^{\tilde{a}}_{-}(x_1)}
+\xi^{\tilde{a}\dag}_{+}(x_1)\frac{\delta}{\delta \xi^{\tilde{a}\dag}_{-}(x_1)}
-\eta^{\tilde{a}}_+(x_1)\frac{\delta}{\delta\eta^{\tilde{a}}_-(x_1)} \nn \\
&+2{\phi_+}^{\tilde{a}}(x_1)\frac{\delta}{\delta C^{\tilde{a}}(x_1)}
-C^{\tilde{a}}(x_1)\frac{\delta}{\delta {\phi_-}^{\tilde{a}}(x_1)}
\biggr], \nn \\
 J_{--}=&\int d x_1 \biggl[
\psi^{\tilde{a}}_{-1}(x_1)\frac{\delta}{\delta \psi^{\tilde{a}}_{+1}(x_1)}
+\xi^{\tilde{a}}_{-}(x_1)\frac{\delta}{\delta \xi^{\tilde{a}}_{+}(x_1)}
+\xi^{\tilde{a}\dag}_{-}(x_1)\frac{\delta}{\delta \xi^{\tilde{a}\dag}_{+}(x_1)}
-\eta^{\tilde{a}}_-(x_1)\frac{\delta}{\delta\eta^{\tilde{a}}_+(x_1)} \nn \\
&-2{\phi_-}^{\tilde{a}}(x_1)\frac{\delta}{\delta C^{\tilde{a}}(x_1)}
+C^{\tilde{a}}(x_1)\frac{\delta}{\delta {\phi_+}^{\tilde{a}}(x_1)}
\biggr], \nn \\
 J_{0}=&\int d x_1 \biggl[
\psi^{\tilde{a}}_{+1}(x_1)\frac{\delta}{\delta\psi^{\tilde{a}}_{+1}(x_1)}
-\psi^{\tilde{a}}_{-1}(x_1)\frac{\delta}{\delta\psi^{\tilde{a}}_{-1}(x_1)} 
\nn \\
&+\xi^{\tilde{a}}_{+}(x_1)\frac{\delta}{\delta\xi^{\tilde{a}}_{+}(x_1)}
+\xi^{\tilde{a}\dag}_{+}(x_1)\frac{\delta}{\delta\xi^{\tilde{a}\dag}_{+}(x_1)}
-\xi^{\tilde{a}}_{-}(x_1)\frac{\delta}{\delta\xi^{\tilde{a}}_{-}(x_1)}
-\xi^{\tilde{a}\dag}_{-}(x_1)\frac{\delta}{\delta\xi^{\tilde{a}\dag}_{-}(x_1)}
\nn \\
&+\eta^{\tilde{a}}_+(x_1)\frac{\delta}{\delta \eta^{\tilde{a}}_+(x_1)}
-\eta^{\tilde{a}}_-(x_1)\frac{\delta}{\delta \eta^{\tilde{a}}_-(x_1)} 
+2{\phi_+}^{\tilde{a}}(x_1)\frac{\delta}{\delta{\phi_+}^{\tilde{a}}(x_1)}
-2{\phi_-}^{\tilde{a}}(x_1)\frac{\delta}{\delta{\phi_-}^{\tilde{a}}(x_1)}
\biggr]. 
\label{Eq:SU2-generators}
\end{align}
The $Q_\pm$ satisfy the following nilpotency relations,%
\begin{align}
 Q_+^{2}&= \left(\text{infinitesimal gauge transformation with parameter
 $\phi_+$}\right)+\frac{M}{3} J_{++}, \nn \\
 Q_-^{2}&= \left(\text{infinitesimal gauge transformation with parameter
 $-\phi_-$}\right)-\frac{M}{3} J_{--}, \nn \\
 \left\{Q_+,Q_-\right\} &= 
\left(\text{infinitesimal gauge transformation with parameter
 $C$}\right)-\frac{M}{3} J_0, 
\label{nil-2}
\end{align}
which satisfy the $SU(2)$ algebra, 
\begin{equation}
 [J_0, J_{\pm\pm}]=\pm2 J_{\pm\pm}, \quad
 [J_{++},J_{--}] = J_0. 
\end{equation}
We see that $(\psi_{+1},\psi_{-1})$, 
$(\xi_+,\xi_-)$, 
$(\xi_+^\dag,\xi_-^{\dag})$, 
$(\eta_+, -\eta_-)$ and $(Q_+, Q_-)$ transform as doublets 
and $(\phi_+, C, -\phi_-)$ as a triplet under the $SU(2)_R$ 
transformation. 

Using $Q_{\pm}$, we can write down the mother theory action as
\footnote{This kind of deformation is discussed for various supersymmetric
Yang-Mills models in \cite{Kato:2011yh}.}
\begin{align}
 S^{mat} = \left(Q_+Q_- - \frac{M}{3}\right){\cal F}^{mat}, 
\label{BTFT form 2}
\end{align}
where ${\cal F}^{mat}$ is 
\begin{equation}
{\cal F}^{mat} = \frac{1}{g^2}\int dx_1 \Tr \left[
 Y D_1 Y^{\dag} -\psi_{+1}\psi_{-1}
-\frac{1}{2}
\xi^\dag_{+}\xi_{-}
-\frac{1}{2}
\xi_{+}\xi_{-}^\dag
- \frac{1}{4}\eta_+\eta_-
\right].
\end{equation}
Here 
the gauge invariant quantity
${\cal F}^{mat}$ is an $SU(2)_R$ singlet, 
\begin{equation}
 J_{\pm\pm}{\cal F}^{mat} = J_0 {\cal F}^{mat} = 0,  
\label{Eq:invariant-help-1}
\end{equation}
and $SU(2)_R$ transformations on a doublet $(Q_+,Q_-)$ are  
\begin{equation}
 J_{\pm\pm}Q_\mp = Q_\pm, \quad J_0 Q_\pm = \pm Q_\pm. 
\label{Eq:invariant-help-2}
\end{equation}
Then we can see the $Q_{\pm}$ invariance of the action as
\begin{align}
 Q_+ S^{mat} &= Q_+^2Q_- {\cal F}^{mat} -\frac{M}{3} Q_+ {\cal F}^{mat} \nn \\
  &=\frac{M}{3}J_{++}Q_- {\cal F}^{mat} -\frac{M}{3}Q_+{\cal F}^{mat} = 0, \nn \\
 Q_- S^{mat} &= \left(\{Q_+,Q_-\}Q_- - Q_+Q_-^2\right){\cal F}^{mat} 
 -\frac{M}{3} Q_- {\cal F}^{mat} \nn \\
&= -\frac{M}{3}J_0 Q_- {\cal F}^{mat} + \frac{M}{3} Q_+ J_{--} {\cal F}^{mat}
 -\frac{M}{3} Q_- {\cal F}^{mat} = 0. 
\label{Eq:invariant-help-3}
\end{align}
We should note that if a general gauge invariant quantity
$\tilde{\cal F}$ is an $SU(2)_R$ singlet, 
the following quantity 
\begin{equation}
\left(Q_+Q_- - \frac{M}{3}\right)\tilde{\cal F}, 
\end{equation}
is always $Q_{\pm}$ invariant.

The action (\ref{Def:1d mother theory}) is obtained
by integrating out the auxiliary fields $\tH_1, H_y, H_y^\dag$,
and it is written by complexified fields as
\begin{equation}
S^{mat} = \int dx_1 \, \left({\cal L}_B^{mat} + {\cal L}_F^{mat}\right),
\end{equation}
\begin{align}
{\cal L}_B^{mat} = \frac{1}{g^2} \Tr\biggl[&
D_1Y D_1Y^\dag
+\frac{1}{4}\left(D_{1}C \right)^2
+D_{1}\phi_{+} D_{1}\phi_{-}
\nn \\
&+\frac{1}{4}[\phi_{+}, \phi_{-}]^2 
+\frac{1}{4}[C, \phi_{+}][\phi_{-}, C] 
+\frac{1}{2}[\phi_{+}, Y] [Y^{\dag}, \phi_{-} ] 
+\frac{1}{2}[\phi_{-}, Y ][Y^{\dag}, \phi_{+}] 
\nn \\
&+\frac{1}{4}[C, Y] [Y^{\dag}, C]  
+\frac{1}{4}[Y, Y^{\dag}]^2
\nn \\
&-\frac{M}{3}Y
D_1Y^{\dag} 
-\frac{M}{2}C
[\phi_{+}, \phi_{-}] 
+\left(\frac{M}{3}\right)^2
\left(\frac{1}{4}C^2 + \phi_{+}\phi_{-}
\right)\biggr],
\end{align}
\begin{align}
{\cal L}_F^{mat} = \frac{1}{g^2} \Tr\biggl[&
+\xi_{-}^\dag D_1 \xi_{+} -\xi_{-} D_1 \xi_{+}^\dag 
+i\eta_{+} D_1 \psi_{-1}
+i\eta_{-} D_1 \psi_{+1}
\nn \\
&
+i\xi_{-} [Y^\dag, \psi_{+1}]
-i\xi_{-}^\dag [Y, \psi_{+1}]
-i\xi_{+} [Y^\dag, \psi_{-1}]
+i\xi_{+}^\dag [Y, \psi_{-1}]
\nn \\
&-\frac{1}{2}\eta_{-} [Y, \xi_{+}^\dag]
-\frac{1}{2}\eta_{-} [Y^\dag, \xi_{+}]
-\frac{1}{2}\eta_{+} [Y, \xi_{-}^\dag]
-\frac{1}{2}\eta_{+} [Y^\dag, \xi_{-}]
\nn \\
&
+ \psi_{-1}[C, \psi_{+1}]
- \psi_{-1}[\phi_{+}, \psi_{-1}]
+ \psi_{+1}[\phi_{-}, \psi_{+1}]
\nn \\
&
- \frac{1}{4}\eta_{+}[C, \eta_{-}]
- \frac{1}{4}\eta_{-}[\phi_{+}, \eta_{-}]
+ \frac{1}{4}\eta_{+}[\phi_{-}, \eta_{+}]
\nn \\
&
-\frac{1}{2}\xi_{-}^\dag [\phi_{+}, \xi_{-}]
-\frac{1}{2}\xi_{-} [\phi_{+}, \xi_{-}^\dag]
+\frac{1}{2}\xi_{+}^\dag [\phi_{-}, \xi_{+}]
+\frac{1}{2}\xi_{+} [\phi_{-}, \xi_{+}^\dag]
\nn \\
&+\frac{1}{2}\xi_{-}^\dag [C, \xi_{+}]
+\frac{1}{2}\xi_{-} [C, \xi_{+}^\dag]
\nn \\
&+\frac{2M}{3}\psi_{+1}\psi_{-1}
+\frac{M}{3}\xi_+\xi_-^\dag
+\frac{M}{3}\xi_+^\dag \xi_-
-\frac{M}{6}\eta_+\eta_- 
\biggr].
\end{align}

The mother theory action and 
$Q_{\pm}$
are invariant under the $U(1)_r$ transformation 
which acts on 
generic fields ${\cal O}_{mat}$ as
\begin{equation}
{\cal O}_{mat} \to e^{ir \theta}{\cal O}_{mat} \label{def: u1symmetry}
\end{equation}
where $\theta$ is a transformation parameter and $r$ represents the $r$-charge
of ${\cal O}_{mat}$.
Each field has each $r$-charge 
described in the Table \ref{Tab:1}.
\begin{table}[htbp]
\begin{center}
\begin{tabular}{|c||c|c|c|}
\hline
 & $Y,\xi_{\pm }, H_{y}$ & $Y^{\dagger},\xi^\dagger_{\pm }, 
H^{\dag}_y$ &
$C, \phi_{\pm},\psi_{\pm 1},\tH_1, v^1,\eta_{\pm}$ \\
 \hline
$r$ & $+1$ & $-1$ & $0$ 
\\
\hline
\end{tabular}
\end{center}
\caption{$U(1)_r$ charges}
\label{U(1)r charges}
\label{Tab:1}
\end{table}
We can confirm the compatibility with 
$Q_{\pm}$ by directly checking each transformation law
in (\ref{Qpm-M1}), (\ref{Qpm-M2}). 
For example, let us see $Q_{\pm}$ transformations acting on the 
$r =1$ fields $Y,\xi_{\pm}, H_y$,  
\begin{align}
 Q_\pm Y&=\xi_{\pm }, \quad
 Q_\pm \xi_{\pm }=\pm [\phi_\pm, Y], \quad
 Q_\mp \xi_{\pm }=\frac{1}{2} [C, Y] \mp H_y, \nn \\
 Q_\pm H_y &=
 [\phi_\pm,\xi_{\mp }]
 \pm\frac{1}{2}[Y, \eta_\pm] 
\mp \frac{1}{2}[C,\xi_{\pm }]
+\frac{M}{3}\xi_{\pm }.  
\end{align}
We can see that also the right hand sides 
still keep the same $r$-charge $r =1$ as the ones of $Y,\xi_{\pm}, H_y$.
Thus $Q_{\pm}$ commute with the $U(1)_r$ transformation.
The $U(1)_r$ invariance of the mother theory action 
follows from the neutrality of the ${\cal F}^{mat}$, 
thanks to the $U(1)_r$ invariance of the $Q_{\pm}$. 
By utilizing the $U(1)_r$ symmetry 
(\ref{def: u1symmetry}), we can 
perform the orbifolding \cite{Cohen:2003xe,Cohen:2003qw,Kaplan:2005ta} 
on the mother theory.
The orbifolding will be explained 
in following sections.

\subsubsection{Orbifold projection}
We define the orbifold projection operator $\hat{\Gamma} \in \Z_N$
acting on a mother theory field ${\cal O}_{mat}^{(\bar{r})}$ 
with $r$-charge $r = \bar{r}$ 
as
\begin{equation}
\hat{\Gamma}({\cal O}_{mat}^{(\bar{r})}) 
= e^{i \frac{2 \pi \bar{r}}{N}} {\cal C} 
({\cal O}_{mat}^{(\bar{r})}) {\cal C}^{-1},
\end{equation}
where the ${\cal C} \in \Z_N \subset U(mN)$ is 
a tensor product of the $N \times N$ matrix and $m \times m$ matrix 
as
\begin{equation}
{\cal C} = \Omega \otimes 1_m.
\end{equation}
Here $\Omega$ is the $N \times N$ diagonal matrix 
\begin{equation}
\Omega = \diag ( e^{i \frac{2 \pi }{N}}, e^{i \frac{4 \pi }{N}}, \ldots, e^{i \frac{2 N \pi }{N}} ) .
\end{equation}
Now we decompose $mN \times mN$ 
matrix elements into $N \times N$ blocks of the 
$m \times m$ submatrices. Indices of 
an $mN \times mN$ matrix valued field $({\cal O}_{mat}^{(\bar{r})})$ are 
represented as
$({\cal O}_{mat}^{(\bar{r})})_{\hat{a}n_2, \hat{b}n'_2}$,
where 
$\hat{a},\hat{b}$ represent the $m\times m$ 
parts and the $n_2, n'_2$ represent the $N\times N$ parts.

The orbifold projection is removing matrix components except the ones
satisfying
\begin{equation}
\left(\hat{\Gamma}({\cal O}_{mat}^{(\bar{r})})\right)_{\hat{a}n_2, \hat{b}n'_2}
= ({\cal O}_{mat}^{(\bar{r})})_{\hat{a}n_2, \hat{b}n'_2}. \label{Def:orbifold}
\end{equation}
We perform this projection on the all fields of the mother theory.
By replacing every mother theory field by its projected field,
we obtain the orbifold lattice gauge theory action.
Under the projection, an $mN \times mN$ matrix is reduced 
to $N$ sets of 
$m \times m$ matrices.
For instance, 
among $mN \times mN$ matrix components
$({\cal O}_{mat}^{(\bar{r})})_{\hat{a}n_2, \hat{b}n'_2}$ 
only the ones with $n'_2 = n_2 + \bar{r}$ can remain as non-zero,
and they are described by the $N$ non-zero
$m \times m$ blocks as ${\cal O}_{n_2}^{(\bar{r})}$ whose matrix components are
$({\cal O}_{n_2}^{(\bar{r})})_{\hat{a}\hat{b}} 
= ({\cal O}_{mat}^{(\bar{r})})_{\hat{a}n_2, \hat{b}n_2+\bar{r}}$.\footnote{We 
can also choose the definition 
$({\cal O}_{n_2'}^{(\bar{r})})_{\hat{a}\hat{b}} = ({\cal O}_{mat}^{(\bar{r})})_{\hat{a}n'_2-\bar{r}, \hat{b}n'_2}$. For the $r = -1$ fields 
$Y^\dag, \xi_{\pm}^\dag, H_y^\dag$, we employ this choice.}
After the projection, indices $n_2$ are regarded as the 
label of $N$ sites in the $N$ periodic lattice. And 
${\cal O}_{n_2}^{(\bar{r})}$ is interpreted as the $m \times m$ matrix valued 
link field pointing from the site $n_2$ to $n_2 + \bar{r}$.
So, for the $r =0$ fields, $N$ non-zero $m \times m$ matrices
$C_{n_2}, \psi_{\pm 1; n_2}, \eta_{\pm; n_2}, 
\phi_{\pm;n_2}, \tH_{1;n_2}, v^1_{n_2}$ become the site fields 
on the $n_2$.
And for the $r = 1$ fields, 
$Y_{n_2}, 
\xi_{\pm,n_2}, H_{y,n_2}$ are link fields pointing 
from $n_2$ to $n_2+1$. (Here their Hermitian conjugates $Y_{n_2}^\dag, 
\xi_{\pm,n_2}^\dag, H_{y,n_2}^\dag$ 
are regarded as the link fields from 
$n_2 +1 $ to $n_2$.)
The gauge symmetry of the mother theory $U(mN)$ is broken to the $U(m)^N$
by the orbifold projection.

After the projection, 
among the 8 supercharges of the mother theory (\ref{dyn_SUSY_2dE}),
only 4 supercharges with $r = 0$ can be preserved on the lattice.
The $Q_{\pm}$ are 2 of the 4 preserved supercharges on the lattice.
The 4 supercharges are associated 
with the fermions on the sites.
Among the 8 SUSY parameters $\e^T = {\e'}^T U_8^{-1}$ in 
(\ref{dyn_SUSY_2dE}),
following 4 components become the SUSY parameter on the lattice,
\begin{equation}
{\e'_{orb}}^T = \left(
\varepsilon_+,  0,  0,  \varepsilon_{\alpha},  \varepsilon_{-},  0,  0,  
\varepsilon_{\beta}
\right).
\end{equation}
Please note our orbifold lattice gauge theory can preserve not only
the $Q_{\pm}$ but also other 2 supercharges associated with 
$\varepsilon_{\alpha}, \varepsilon_{\beta}$.
In the lattice formulation in \cite{Hanada:2011qx},
only the $Q_{\pm}$ can be preserved on the lattice.

The orbifolded action can also be written as
\begin{align}
 S^{orb} = \left(Q_+Q_- - \frac{M}{3}\right){\cal F}^{orb}, 
\label{Eq:BTFT form orb}
\end{align}
where ${\cal F}^{orb}$ is 
\begin{equation}
{\cal F}^{orb} = \frac{1}{g^2}\int dx_1 \sum_{n_2}\tr \left[
Y_{n_2} {\cal D}_1 Y^{\dag}_{n_2} -\psi_{+1;n_2}\psi_{-1;n_2}
-\frac{1}{2}
\xi^\dag_{+;n_2}\xi_{-;n_2}
-\frac{1}{2}
\xi_{+;n_2}\xi_{-;n_2}^\dag
- \frac{1}{4}\eta_{+;n_2}\eta_{-;n_2}
\right],
\end{equation}
in the same way as 
(\ref{BTFT form 2}). 
Here the "$\tr$" denotes trace over the $m \times m$ matrix.
The covariant derivative ${\cal D}_1$ is defined for 
link fields as
\begin{equation}
{\cal D}_1 Y_{n_2} \equiv \del_1 Y_{n_2}^\dag
+i v^1_{n_2 }Y_{n_2} -i Y_{n_2} v^1_{n_2 +1},
\qquad
{\cal D}_1 Y^{\dag}_{n_2} \equiv \del_1 Y_{n_2}^\dag
+i v^1_{n_2 +1}Y_{n_2}^\dag -i Y_{n_2}^\dag v^1_{n_2 }.
\end{equation}
After the orbifold projection, the $Q_{\pm}$ transformations are written as
\begin{align}
 Q_\pm v^1_{n_2}&=\psi_{\pm 1;n_2}, \quad
 Q_\pm \psi_{\pm 1;n_2}=\pm iD_1 \phi_{\pm;n_2}, \quad
 Q_\mp \psi_{\pm 1;n_2}=\frac{i}{2} D_1 C_{n_2} \mp \tH_{1;n_2}, \nn \\
 Q_\pm \tH_{1;n_2} &=
 [\phi_{\pm;n_2},\psi_{\mp 1;n_2}]\mp \frac{1}{2}[C_{n_2},\psi_{\pm 1;n_2}]
 \mp\frac{i}{2}D_1 \eta_{\pm;n_2} +\frac{M}{3}\psi_{\pm 1;n_2},  \nn \\
 Q_\pm Y_{n_2}&=\xi_{\pm ;n_2}, \quad
 Q_\pm \xi_{\pm;n_2 }=\pm \left(\phi_{\pm;n_2} Y_{n_2} - Y_{n_2}
\phi_{\pm;n_2+1}\right), 
\nn \\
 Q_\mp \xi_{\pm;n_2 }&=\frac{1}{2} \left(C_{n_2}Y_{n_2} - Y_{n_2} C_{n_2+1}
\right) \mp H_{y;n_2}, \nn \\
 Q_\pm H_{y;n_2} &=
 \left(
\phi_{\pm;n_2}\xi_{\mp;n_2 }
-\xi_{\mp;n_2 }\phi_{\pm;n_2+1}
\right)
\mp \frac{1}{2}\left(
C_{n_2}\xi_{\pm;n_2 }
-\xi_{\pm;n_2 }C_{n_2+1}
\right) \nn \\
& \quad \pm\frac{1}{2}\left(
Y_{n_2} \eta_{\pm;n_2+1}
- \eta_{\pm;n_2}Y_{n_2}
\right)
 +\frac{M}{3}\xi_{\pm;n_2 },  \nn \\
 Q_\pm Y^\dag_{n_2} &= \xi_{\pm;n_2}^\dag, \quad 
 Q_\pm \xi_{\pm;n_2}^\dag = \pm\left(
\phi_{\pm; n_2 +1} Y^{\dag}_{n_2}
-Y^{\dag}_{n_2}\phi_{\pm; n_2 } 
\right), 
\nn \\
 Q_\mp \xi_{\pm;n_2 }^\dag&=\frac{1}{2} 
\left(
C_{n_2+1} Y^\dag_{n_2}
-Y^\dag_{n_2}C_{n_2} 
\right) \mp H_{y;n_2}^\dag, \nn \\
 Q_\pm H_{y;n_2}^\dag &=
 \left(
\phi_{\pm;n_2+1}\xi_{\mp;n_2 }^\dag
-\xi_{\mp;n_2 }^\dag\phi_{\pm;n_2}
\right)
\mp \frac{1}{2}
\left(
C_{n_2+1}\xi_{\pm;n_2 }^\dag
-\xi_{\pm;n_2 }^\dag C_{n_2}
\right)
\nn \\
& \quad \pm\frac{1}{2}\left(
Y^\dag_{n_2} \eta_{\pm;n_2}
-\eta_{\pm;n_2+1}Y^\dag_{n_2} 
\right) 
+\frac{M}{3}\xi_{\pm;n_2 }^\dag,  \nn \\
 Q_\pm C_{n_2} &= \eta_{\pm;n_2}, \quad
 Q_\pm \phi_{\pm;n_2} = 0, \quad
 Q_\mp \phi_{\pm;n_2} = \mp \eta_{\pm;n_2}, \nn \\
 Q_\pm \eta_{\pm;n_2} &= \pm[\phi_{\pm;n_2},C_{n_2}]+\frac{2M}{3}
\phi_{\pm;n_2}, \quad
 Q_\mp \eta_{\pm;n_2} 
= \mp[\phi_{+;n_2},\phi_{-;n_2}]\pm\frac{M}{3}C_{n_2}. 
 \label{Qpm-M}
\end{align}

\subsubsection{Tree level continuum limit}
For the discussion of the moduli fixing terms in the next sub-subsection,
we will consider the tree level continuum limit.
To see the tree level continuum limit,
we perform the deconstruction, which is expanding the 
bosonic link fields around $\frac{1}{a}$,
\begin{equation}
Y_{n_2} = \frac{1}{a} + s_{3;n_2} + i v_{2;n_2},
\qquad
Y_{n_2}^\dag = \frac{1}{a} + s_{3;n_2} - i v_{2;n_2},
\label{Eq:deconstruction}
\end{equation}
where the $a$ is regarded as lattice spacing.
We also identify the site fields as
\begin{equation}
\phi_{+;n_2} = s^5_{n_2} + is^6_{n_2}, \quad \phi_{-;n_2}
= s^5_{n_2} -i s^6_{n_2}, \qquad 
C_{n_2} = s^4_{n_2}, 
\end{equation}
where the $s^i \,(i =3,4,5,6)$ are the scalar fields
in the two-dimensional super Yang-Mills theory with 8 supercharges.
By substituting the above into the orbifold lattice action
(\ref{Eq:BTFT form orb})
and performing the Taylor expansion with respect to the lattice spacing 
$a$ around $a =0$, 
we can see the target continuum limit at the tree level.
During the procedure, we interpret $g^2 a = g_{2d}^2$ as the 
two-dimensional gauge coupling and it is fixed under the limit $a \to 0$.
The target continuum limit $(a \to 0)$ is 
\begin{align}
S_{2d,0} = \frac{2}{g_{2d}^2}\int d^2 x \, \tr \biggl[& 
\frac{1}{2}F_{12}^2
+\frac{1}{2}D_{\mu}s^iD_{\mu}s^i
-\frac{1}{4}[s^i, s^j]^2 
+\frac{i}{2}\Psi^T \gamma^{i} [s_i, \Psi]
+\frac{1}{2}\Psi^T (D_1 + \gamma_2 D_2) \Psi
\nn \\
&
-i\frac{M}{6}{\Psi}^T\gamma^{23}{\Psi} 
+\frac{iM}{3}{s}^3F_{12} +\frac12\left(\frac{M}{3}\right)^2 ({s}^a)^2
 +i\frac{M}{3}\epsilon_{abc}{s}^a{s}^b{s}^c
\biggr],\nn \\ 
\label{Eq: 2d continuum theory}
\end{align}
where the subscripts $\mu$ run $1,2$ and $i,j = 3,4,5,6$. 
The field $v_2$ is regarded as the gauge field along the $x_2$ direction and
the $F_{12}$ is the gauge field strength, 
$F_{12} = \del_1 v^2 - \del_2 v^1 +i [v^1,v^2]$.
Fermionic fields on the lattice are recombined to be the two-dimensional 
8 component spinor $\Psi$ in the same way as (\ref{Fermion reps})-(\ref{Uf}),
and the $\xi_{\pm}$ are reinterpreted as
$\xi_{\pm}  =  i \psi_{\pm 2} + \chi_{\pm}$, 
$\xi_{\pm}^\dag  = - i \psi_{\pm 2} + \chi_{\pm}$.
The continuum action (\ref{Eq: 2d continuum theory})
is the mass-deformed two-dimensional supersymmetric Yang-Mills theory,
which is same as (3.7) or (A.28) in the \cite{Hanada:2011qx}.

\subsubsection{Stabilization of the vacuum without breaking lattice SUSY}
It is necessary to justify the expansion 
around the point
\begin{equation}
Y_{n_2} = Y^{\dag}_{n_2} = 1/a
\end{equation}
to make a lattice interpretation of our action 
(\ref{Eq:BTFT form orb}) with the target continuum limit
(\ref{Eq: 2d continuum theory}).
In order to justify, 
quantum fluctuation around the classical vacua must be
small enough compared to the classical value $1/a$.
Actually, as pointed out also in \cite{Hanada:2011qx},
there is a flat direction along $s_3$ 
which allows the 
large fluctuation spoiling the lattice interpretation.
In a conventional way to lift the degeneracy of the vacua, 
soft SUSY breaking mass terms might be introduced.
Although such mass terms do not alter UV divergences, they break the 
supersymmetry on the lattice.

In this paper, as an alternative way, 
we would like to propose new moduli fixing 
terms which will {\it not} break the supersymmetry on the lattice,
\begin{equation}
S_{mass}^{orb} 
= \frac{a^2 \nu_{1}}{g^2} \left(Q_{+}Q_{-} -\frac{1}{3}M\right)
\int dx_1 \, \sum_{n_2}
\tr \left( 
Y_{n_2} Y^\dag_{n_2} - \frac{1}{a^2}
\right)^2,
\label{SUSY preserving mass term}
\end{equation}
where the $\nu_1$ is a mass parameter with mass dimension 1.
These are very analogous to the mass terms of the $B(x)$ in the 
\cite{Hanada:2011qx}.
We should note that $Y_{n_2},Y^{\dag}_{n_2}$ are 
singlet under the $SU(2)_R$ symmetry generated by
(\ref{Eq:SU2-generators}), and the lattice spacing $a$ is 
a non-dynamical quantity 
vanishing under the $Q_{\pm}$
transformation. 
Of course $a$ itself is invariant under the $SU(2)_R$.
Since 
$\tr \left( 
Y_{n_2} Y^\dag_{n_2} - \frac{1}{a^2}
\right)^2
$
is a gauge singlet as well as an $SU(2)_R$ singlet,
we can see that the mass terms (\ref{SUSY preserving mass term})
are invariant under the $Q_{\pm}$
from 
(\ref{Eq:invariant-help-1})-
(\ref{Eq:invariant-help-3}). 
Namely $Q_{\pm}$ are still preserved 
on the lattice in presence of the new moduli fixing terms
(\ref{SUSY preserving mass term}).

The fixing terms 
(\ref{SUSY preserving mass term})
include the auxiliary fields.
If we integrate out the 
auxiliary fields
after summing up the 
(\ref{Eq:BTFT form orb}) and (\ref{SUSY preserving mass term}), 
the terms depending on the
$\nu_1$ become,
\begin{align}
S_{mass}^{orb} 
= \frac{1}{g^2} 
\int dx_1 \, \sum_{n_2} \tr \biggl[
&2a^2 \nu_{1}\left( -\xi_{-;n_2} \xi_{+;n_2}^{\dag} 
+ \xi_{+;n_2} \xi_{-;n_2}^{\dag} 
\right){\cal Y}_{n_2}
\nn \\
&-2a^2 \nu_{1}\left( \xi_{-;n_2} Y_{n_2}^{\dag} 
+ Y_{n_2} \xi_{-;n_2}^{\dag} 
\right)\left( \xi_{+;n_2} Y_{n_2}^{\dag} 
+ Y_{n_2} \xi_{+;n_2}^{\dag} 
\right)
\nn \\
&+2a^2 \nu_{1}\left( 
{\cal D}_1 Y_{n_2} Y^\dag_{n_2}{\cal Y}_{n_2}
-{\cal Y}_{n_2} Y_{n_2} {\cal D}_1 Y^\dag_{n_2}
\right)
\nn \\
&-\frac{M a^2 \nu_{1}}{3}
{\cal Y}_{n_2}^2
-4 a^4 \nu_{1}^2 Y_{n_2} Y^\dag_{n_2} 
{\cal Y}_{n_2}
{\cal Y}_{n_2}
\biggr], \label{SUSY preserving mass term explicit 2}
\end{align}
where ${\cal Y}_{n_2} = Y_{n_2} Y^\dag_{n_2} - \frac{1}{a^2}$. 
At the continuum limit, these mass terms become
\begin{equation}
S_{mass}^{orb} \,\,
\stackrel{a \to 0}{\longrightarrow}
\, 
\frac{1}{g_{2d}^2}\int d^2 x \, \tr \left(
\left(-16\nu_1^2 - \frac{4M\nu_1}{3}\right) s_3^2
+8i\nu_1
s_3 F_{12}
-8 \nu_1 \chi_{-}\chi_{+}
\right),
\end{equation}
these are the same as
the eq.~(3.12) in the \cite{Hanada:2011qx}.
Here each $\chi_{+}$ and $\chi_{-}$ is a spinor component in the $\Psi$
written in (\ref{Fermion reps})-(\ref{Uf}).

In our case, we also have to take care of the 
IR divergence $\sim g_2^2 \log (a\tilde{\nu})$ where the
$\tilde{\nu}$ is the IR cut-off.
To keep the divergence much smaller than the classical values,
$a^2 g_2^2\log(a\tilde{\nu}) \ll 1$ must be satisfied.
In order to take the continuum limit with 
keeping $a^2 g_2^2\log(a\tilde{\nu}) \ll 1$,
we need to separate the procedure of taking the limits 
into the following two steps:
\begin{itemize}
\item In the orbifold lattice theory, first we should take
$a \to 0$ and $N \to \infty$,
with keeping $\nu_1$ (or $\bar{\nu}_1, \bar{\nu}_2$) and $aN$ fixed.
\item After that, $\nu_1 \to 0$ as well as $aN \to \infty$ 
(or $\bar{\nu}_1, \bar{\nu}_2 \to 0$) 
to recover the full 8 supercharges. 
\end{itemize}
In this way, the step 2 and the step 3,
which are a part of 
the steps of taking the limits explained in Sec.~\ref{Sec:steps}, 
get separated from each other.

There is also another constraint on the parameter $\nu_1$.
In order for $s_3$ to have a positive mass squared,
$\nu_1$ must satisfy
\begin{equation}
-\frac{M}{12} < \nu_1 < 0.
\end{equation}
So in presence of the SUSY preserving moduli fixing terms
(\ref{SUSY preserving mass term}),
we also have to be careful about parameter region of the 
$\nu_1$.\footnote{This mass term will provide not only the mass term
for the $SU(m)$ part, but also the $U(1)$ part decoupling from the 
$SU(m)$ part.}

For a practical usage, if one prefers a simpler treatment 
which does not include fermionic terms,
also adding conventional soft breaking mass terms
\begin{equation}
\tilde{S}_{mass}^{orb} 
= \frac{1}{g^2}\int dx_1 \, \sum_{n_2} \left(
a^2 \bar{\nu}_{1}^2 \,\,\tr \left( 
Y_{n_2} Y^\dag_{n_2} - \frac{1}{a^2}
\right)^2
+ a^2 \bar{\nu}_2^2 \left| \tr(Y_{n_2}Y^\dag_{n_2}) - \frac{1}{a^2} \right|^2
\right) \label{soft breaking mass term}
\end{equation}
could work.
Because these moduli fixing terms are just soft SUSY breaking terms,
they would not alter the UV divergences.
But conclusions of the numerical calculation would be 
more or less obscured since the soft breaking terms break the lattice SUSY.
The SUSY preserving fixing terms 
(\ref{SUSY preserving mass term})
would help to 
get more concrete conclusions. 

\subsubsection{Orbifold lattice action with the moduli fixing terms}
Finally, by adding the moduli fixing terms 
(\ref{SUSY preserving mass term explicit 2}), 
we complete the construction of the orbifold lattice action as
\begin{equation}
S^{orb} = S^{orb}_B + S^{orb}_F,
\label{def:orbifold action}
\end{equation}
where the $S^{orb}_B$ is the bosonic part described as
\begin{align}
S_B^{orb} 
= \frac{1}{g^2}\int dx_1 \sum_{n_2}\tr  \biggl[&
(\del_{x_1}Y_{n_2} + iv^1_{n_2}Y_{n_2} - i Y_{n_2}v^1_{n_2+1})
(\del_{x_1}Y_{n_2}^{\dag} 
+ iv^1_{n_2+1}Y_{n_2}^{\dag} - i Y_{n_2}^{\dag}v^1_{n_2})
\nn \\
&+\frac{1}{4}\left(D_{1}C_{n_2} \right)^2
+D_{1}\phi_{+;n_2} D_{1}\phi_{-;n_2}
+\frac{1}{4}[\phi_{+;n_2}, \phi_{-;n_2}]^2 
+\frac{1}{4}[C_{n_2}, \phi_{+;n_2}][\phi_{-;n_2}, C_{n_2}] 
\nn \\
&+
\frac{1}{2}(\phi_{+;n_2} Y_{n_2} - Y_{n_2} \phi_{+;n_2+1}) 
(Y^{\dag}_{n_2} \phi_{-;n_2}  -\phi_{-;n_2+1} Y^{\dag}_{n_2}) 
\nn \\
&+\frac{1}{2}(\phi_{-;n_2} Y_{n_2} - Y_{n_2} \phi_{-;n_2+1}) 
(Y^{\dag}_{n_2} \phi_{+;n_2}  -\phi_{+;n_2+1} Y^{\dag}_{n_2}) 
\nn \\
&+\frac{1}{4}(C_{n_2} Y_{n_2} - Y_{n_2} C_{n_2+1}) 
(Y^{\dag}_{n_2} C_{n_2}  - C_{n_2+1} Y^{\dag}_{n_2}) 
\nn \\
&-\frac{1}{4}
(Y_{n_2} Y^{\dag}_{n_2} - Y^{\dag}_{n_2-1} Y_{n_2-1})      
(Y^{\dag}_{n_2-1} Y_{n_2-1} - Y_{n_2} Y^{\dag}_{n_2})
\nn \\
&-\frac{M}{3}\left(Y_{n_2}\del_{x_1} 
Y^{\dag}_{n_2} 
+iY_{n_2}v^1_{n_2+1}Y^{\dag}_{n_2}
-iY_{n_2}Y^{\dag}_{n_2}v^1_{n_2} \right)
\nn \\
&-\frac{M}{2}C_{n_2}
[\phi_{+;n_2}, \phi_{-;n_2}] 
+\left(\frac{M}{3}\right)^2
\left(\frac{1}{4}C_{n_2}^2 + \phi_{+;n_2}\phi_{-;n_2}
\right)
\nn \\
&+2a^2 \nu_{1}\left( 
{\cal D}_1 Y_{n_2} Y^\dag_{n_2}{\cal Y}_{n_2}
-{\cal Y}_{n_2} Y_{n_2} {\cal D}_1 Y^\dag_{n_2}
\right)
\nn \\
&-\frac{M a^2 \nu_{1}}{3}
{\cal Y}_{n_2}^2
-4 a^4 \nu_{1}^2 Y_{n_2} Y^\dag_{n_2} 
{\cal Y}_{n_2}
{\cal Y}_{n_2}
\biggr].
\label{def:orbifold action boson}
\end{align}
The fermionic part $S^{orb}_F$ is
\be
S_F^{orb}
= \int dx^1 \left( {\cal L}_{F1}^{orb}+{\cal L}_{F2}^{orb}
+{\cal L}_{F3}^{orb}+ {\cal L}_{mF}^{orb} \right),
\label{def:orbifold fermion}
\ee
where
\begin{align}
{\cal L}_{F1}^{orb}= \frac{1}{g^2}
\sum_{n_2} \tr  \biggl[&
+i\eta_{+;n_2} D_1 \psi_{-1;n_2}
+i\eta_{-;n_2} D_1 \psi_{+1;n_2}
+\xi_{-;n_2}^\dag \del_1 \xi_{+;n_2} -\xi_{-;n_2} \del_1 \xi_{+;n_2}^\dag 
\nn \\
&+i\xi_{-;n_2}^\dag 
(v_{1;n_2} \xi_{+;n_2} -  \xi_{+;n_2}v_{1;n_2+1})
-i\xi_{-;n_2} 
(v_{1;n_2+1} \xi_{+;n_2}^\dag -  \xi_{+;n_2}^\dag v_{1;n_2})\nn \\
&
+i\xi_{-;n_2} \left(
Y^\dag_{n_2} \psi_{+1;n_2} -\psi_{+1;n_2+1}Y^\dag_{n_2} 
\right)
-i\xi_{-;n_2}^\dag 
\left(
Y_{n_2} \psi_{+1;n_2+1}- \psi_{+1;n_2} Y_{n_2}
\right)
\nn \\
&-i\xi_{+;n_2} 
\left(
Y^\dag_{n_2} \psi_{-1;n_2}- \psi_{-1;n_2+1}Y^\dag_{n_2}
\right)
+i\xi_{+;n_2}^\dag 
\left(
Y_{n_2} \psi_{-1;n_2+1} -\psi_{-1;n_2} Y_{n_2}
\right) \biggr],
\end{align}
\begin{align}
{\cal L}_{F2}^{orb}= \frac{1}{g^2}
\sum_{n_2} \tr  \biggl[
&-\frac{1}{2}
\left(\eta_{-;n_2} Y_{n_2} - Y_{n_2} \eta_{-;n_2+1}\right)
\xi_{+;n_2}^\dag
-\frac{1}{2}
\left(
\eta_{-;n_2+1} Y^\dag_{n_2}
- Y^\dag_{n_2}\eta_{-;n_2}
\right) \xi_{+;n_2}
\nn \\
&-\frac{1}{2}\left(
\eta_{+;n_2} Y_{n_2}-Y_{n_2} \eta_{+;n_2+1} 
\right) \xi_{-;n_2}^\dag
-\frac{1}{2}
\left(
\eta_{+;n_2+1} Y^\dag_{n_2}- Y^\dag_{n_2}\eta_{+;n_2}
\right) \xi_{-;n_2}
\nn \\
&
+ \psi_{-1;n_2}[C_{n_2}, \psi_{+1;n_2}]
- \psi_{-1;n_2}[\phi_{+;n_2}, \psi_{-1;n_2}]
+ \psi_{+1;n_2}[\phi_{-;n_2}, \psi_{+1;n_2}]
\nn \\
&
- \frac{1}{4}\eta_{+;n_2}[C_{n_2}, \eta_{-;n_2}]
- \frac{1}{4}\eta_{-;n_2}[\phi_{+;n_2}, \eta_{-;n_2}]
+ \frac{1}{4}\eta_{+;n_2}[\phi_{-;n_2}, \eta_{+;n_2}]
 \biggr],
\end{align}
\begin{align}
{\cal L}_{F3}^{orb}=
\frac{1}{ g^2} 
 \sum_{n_2} \tr \biggl[&
-\frac{1}{2}\xi_{-;n_2}^\dag 
\left(
\phi_{+;n_2}\xi_{-;n_2}-\xi_{-;n_2}\phi_{+;n_2+1}
\right)
-\frac{1}{2}\xi_{-;n_2} 
\left(
\phi_{+;n_2+1} \xi_{-;n_2}^\dag- \xi_{-;n_2}^\dag \phi_{+;n_2}
\right)
\nn \\
&+\frac{1}{2}\xi_{+;n_2}^\dag 
\left(
\phi_{-;n_2} \xi_{+;n_2}- \xi_{+;n_2}\phi_{-;n_2+1}
\right)
+\frac{1}{2}\xi_{+;n_2} 
\left(
\phi_{-;n_2+1} \xi_{+;n_2}^\dag- \xi_{+;n_2}^\dag\phi_{-;n_2}
\right)
\nn \\
&+\frac{1}{2}\xi_{-;n_2}^\dag 
\left(
C_{n_2} \xi_{+;n_2}- \xi_{+;n_2}C_{n_2+1}
\right)
+\frac{1}{2}\xi_{-;n_2} 
\left(
C_{n_2+1} \xi_{+;n_2}^\dag- \xi_{+;n_2}^\dag C_{n_2}
\right) \biggr],
\end{align}
\begin{align}
{\cal L}_{mf}^{orb} =
 \frac{1}{g^2}\sum_{n_2} \tr \biggl[&
\frac{2M}{3}\psi_{+1;n_2}\psi_{-1;n_2}
+\frac{M}{3}\xi_{+;n_2}\xi_{-;n_2}^\dag
+\frac{M}{3}\xi_{+;n_2}^\dag \xi_{-;n_2}
-\frac{M}{6}\eta_{+;n_2}\eta_{-;n_2} 
\nn \\
&+2a^2 \nu_{1}\left( -\xi_{-;n_2} \xi_{+;n_2}^{\dag} 
+ \xi_{+;n_2} \xi_{-;n_2}^{\dag} 
\right){\cal Y}_{n_2}
\nn \\
&-2a^2 \nu_{1}\left( \xi_{-;n_2} Y_{n_2}^{\dag} 
+ Y_{n_2} \xi_{-;n_2}^{\dag} 
\right)\left( \xi_{+;n_2} Y_{n_2}^{\dag} 
+ Y_{n_2} \xi_{+;n_2}^{\dag} 
\right)
\biggr].
\end{align}

\subsection{Momentum cut-off regularization on the orbifold lattice gauge theory (the hybrid regularization theory)}
\label{Momentum cut-off}
To perform numerical studies, 
we need to regularize the continuum $x_1$ direction 
of the orbifold lattice theory (\ref{def:orbifold action})
also. Here $x_1$ is periodic $x_1 \sim x_1 + R_1$, and we
impose the periodic boundary condition on all fields.

At finite $mN$, 
the orbifold lattice gauge theory can be regarded as a
one-dimensional system.
Since the orbifold projection is 
just picking up a part of the $mN \times mN$ matrices,
the action 
can be embedded into a one-dimensional $mN \times mN$ matrix model system.
As we explain in a later section Sec.~\ref{sec:1d correction},
UV divergences are usually absent in one-dimensional systems.
Then even if we regularize a one-dimensional system in a naive way,
it is easy to recover the target continuum limit 
without any fine-tunings.
Here we apply the momentum cut-off regularization implemented in the 
\cite{Hanada:2007ti,Anagnostopoulos:2007fw,Hanada:2008gy,Hanada:2008ez,Hanada:2009ne,Hanada:2010jr,Hanada:2011fq}.\footnote{There is also another choice
to employ the lattice regularization 
like in \cite{Catterall:2000rv}.} 

To implement the momentum cut-off regularization, 
first we need to 
gauge-fix the $U(m)^N$ gauge symmetry. 
Details of the gauge fixing are described in the appendix
\ref{Sec:gauge fixing}.
Here we will impose 
the static diagonal gauge, 
\begin{equation}
v^1_{n_2}(x_1) 
= \frac{1}{R_1} \diag (\alpha_{1n_2}, \ldots,\alpha_{\hat{a}n_2}, \ldots 
\alpha_{mn_2}).
\end{equation}
Here the $\alpha_{\hat{a}n_2}$ are dimensionless constants with
respect to $x_1$.
By this gauge fixing, 
we need to include the following Fadeev-Popov determinant term,
\begin{equation}
S_{FP} = - \sum_{\hat{a} <\hat{b}} \sum_{n_2} 2 
\log \left|\sin \frac{\alpha_{\hat{a}n_2} - \alpha_{\hat{b}n_2}}{2}\right|.
\end{equation}
Here the ghost fields are sitting on sites.
We can set the domain of $\alpha_{n_2}^{\hat{a}}$ as  
$\max (\alpha_{n_2}^{\hat{a}}) - \min (\alpha_{n_2}^{\hat{b}}) \le 2 \pi$  
by fixing the residual 
large gauge transformations with non-zero 
winding numbers.\footnote{It 
is necessary to fix the large gauge transformations 
to justify the momentum cut-off.
If they are not fixed, there is a risk to allow the momentum 
to go beyond the cut-off $\Lambda$ since the transformations 
have an effect to shift the momentum.}
Here the integration measure is 
taken to be uniform.\footnote{
In the $U(m)^N$ theory, there are
decoupling $U(1)$ zero mode of site fields. 
It is required to remove them for the simulation, in particular 
we need to remove the fermionic decoupling modes 
to avoid the zero fermion determinant.}
After performing the gauge fixing,
we make a Fourier expansion 
with the UV cut-off $\Lambda$ as
\begin{equation}
\Phi^{\hat{a}\hat{b}}_{n_2}(x_1) = 
\sum_{p_1 = - \Lambda}^{\Lambda}
\hat{\Phi}^{\hat{a}\hat{b}}_{n_2, p_1}
e^{i p_1 \omega x_1},
\end{equation}
where the $\Phi$ denotes general field variables in the 
orbifold lattice gauge theory and 
$\omega = \frac{2\pi}{R_1}$ and $p_1$ is integer.
By substituting the Fourier expansion into the action, we
will obtain the
(1+1)-dimensional hybrid regularization theory action as
\begin{equation}
S^{\Lambda} = S^{\Lambda}_B + S^{\Lambda}_F.
\label{Eq:cut-off action}
\end{equation}
Here the bosonic part $S^{\Lambda}_B$ is 
\begin{equation}
S_B^{\Lambda} = S_{Bp}^{\Lambda} + 
S_{B0}^{\Lambda} 
+S_{Bm}^{\Lambda} + 
S_{B\nu_1 p}^{\Lambda} + 
S_{B\nu_1 0}^{\Lambda}  ,
\end{equation}
where
\begin{align}
S^{\Lambda}_{Bp} = 
\frac{1}{g^2} \sum_{n_2} 
\sum_{\hat{a},\hat{b}}\sum_{p_1 = - \Lambda}^{\Lambda}
\biggl[
&\left(ip_1 \omega 
+ i\frac{\alpha^{\hat{a}}_{n_2} - \alpha^{\hat{b}}_{n_2+1} }{R_1} 
\right)
\left(-ip_1 \omega 
+ i\frac{\alpha^{\hat{b}}_{n_2+1} - \alpha^{\hat{a}}_{n_2}}{R_1}
\right) 
\hat{Y}_{n_2, p_1}^{\hat{a} \hat{b}}
\hat{Y}^{\hat{b} \hat{a}\,\dag}_{n_2, -p_1}
\nn \\
&+\frac{1}{4}
\left(ip_1 \omega+ 
i\frac{\alpha^{\hat{a}}_{n_2} - \alpha^{\hat{b}}_{n_2}}{R_1} \right)
\left(-ip_1 \omega+ i\frac{\alpha^{\hat{b}}_{n_2} 
- \alpha^{\hat{a}}_{n_2}}{R_1} \right)
\hat{C}^{\hat{a}\hat{b}}_{n_2, p_1} 
\hat{C}^{\hat{b}\hat{a}}_{n_2, -p_1} \nn \\
&+\left(ip_1 \omega+ 
i\frac{\alpha^{\hat{a}}_{n_2} - \alpha^{\hat{b}}_{n_2}}{R_1} \right)
\left(-ip_1 \omega+ i\frac{\alpha^{\hat{b}}_{n_2} 
- \alpha^{\hat{a}}_{n_2}}{R_1} \right)
\hat{\phi}^{\hat{a}\hat{b}}_{+;n_2, p_1} 
\hat{\phi}^{\hat{b}\hat{a}}_{-;n_2, -p_1} \nn \\
&+\left(-\frac{M}{3} + 4 \nu_1\right)
\left( -i p_1 \omega + i 
\frac{\alpha^{\hat{b}}_{n_2+1} -  \alpha^{\hat{a}}_{n_2}}{R_1}\right) 
\hat{Y}_{n_2, p_1}^{\hat{a}\hat{b}}
\hat{Y}_{n_2, -p_1}^{\hat{b}\hat{a}\dag}
 \biggr],
\end{align}
\begin{align}
S_{B0}^{\Lambda} = \frac{1}{g^2}\sum_{n_2} \tr
\biggl[&
\frac{1}{4}\left([\hph_{+;n_2}, \hph_{-;n_2}]^2\right)_0 
+\frac{1}{4}\left([\hC_{n_2}, \hph_{+;n_2}][\hph_{-;n_2}, \hC_{n_2}]
\right)_0
\nn \\
&+
\frac{1}{2}\left((\hph_{+;n_2} \hY_{n_2} - \hY_{n_2} \hph_{+;n_2+1}) 
(\hY^{\dag}_{n_2} \hph_{-;n_2}  -\hph_{-;n_2+1} \hY^{\dag}_{n_2})\right)_0
\nn \\
&+\frac{1}{2}\left((\hph_{-;n_2} \hY_{n_2} - \hY_{n_2} \hph_{-;n_2+1}) 
(\hY^{\dag}_{n_2} \hph_{+;n_2}  -\hph_{+;n_2+1} \hY^{\dag}_{n_2})\right)_0
\nn \\
&+\frac{1}{4}\left((\hC_{n_2} \hY_{n_2} - \hY_{n_2} \hC_{n_2+1}) 
(\hY^{\dag}_{n_2} \hC_{n_2}  - \hC_{n_2+1} \hY^{\dag}_{n_2})\right)_0
\nn \\
&-\frac{1}{4}\left(
(\hY_{n_2} \hY^{\dag}_{n_2} - \hY^{\dag}_{n_2-1} \hY_{n_2-1})      
(\hY^{\dag}_{n_2-1} \hY_{n_2-1} - \hY_{n_2} \hY^{\dag}_{n_2})\right)_0
 \biggr],
\end{align}
\begin{align}
S_{Bm}^{\Lambda} = \frac{1}{g^2}\sum_{n_2} \tr
\biggl[&-\frac{M}{2}\left(\hC_{n_2}
[\hph_{+;n_2}, \hph_{-;n_2}] \right)_0
+\left(\frac{M}{3}\right)^2
\left(\frac{1}{4}\left(\hC_{n_2}^2\right)_0 + 
\left(\hph_{+;n_2}\hph_{-;n_2}\right)_0
\right)\biggr],
\end{align}
\begin{align}
S_{B\nu_1 p}^{\Lambda} = \frac{2a^2 \nu_{1}}{g^2} \sum_{n_2} 
\sum_{\hat{a},\hat{b}}\sum_{p_1 = - \Lambda}^{\Lambda}
\biggl[&
\left(ip_1 \omega+ 
i\frac{\alpha^{\hat{a}}_{n_2} - \alpha^{\hat{b}}_{n_2+1}}{R_1} \right)
\hY_{n_2,p_1}^{\hat{a}\hat{b}} 
\left(\hY^\dag_{n_2}\hY_{n_2}\hY^\dag_{n_2}\right)^{\hat{b}\hat{a}}_{-p_1}
\nn \\
&-
\left(ip_1 \omega+ 
i\frac{\alpha^{\hat{a}}_{n_2+1} - \alpha^{\hat{b}}_{n_2}}{R_1} \right)
\hY_{n_2,p_1}^{\dag\hat{a}\hat{b}}
\left(\hY_{n_2}\hY^\dag_{n_2} \hY_{n_2} \right)^{\hat{b}\hat{a}}_{-p_1}
\biggr],
\end{align}
\begin{align}
S_{B\nu_1 0}^{\Lambda} = \frac{1}{g^2} \sum_{n_2} 
\biggl[
&-\frac{M a^2 \nu_{1}}{3}
\left(\hat{\cal Y}_{n_2}^2\right)_0
-4 a^4 \nu_{1}^2 \left(\hY_{n_2} \hY^\dag_{n_2} 
\hat{\cal Y}_{n_2}
\hat{\cal Y}_{n_2}\right)_0
\biggr]. 
\end{align}
Here of course the momentum along the $x_1$ direction 
is conserved at each vertex,
and each subscript ${}_0$ and ${}_{-p_1}$ means 
the sum of the 
all assignments of the momentum such that 
the total momentum is $0$ or $-p_1$ respectively.
The fermionic part of the action $S_{F}^{\Lambda}$ is 
\begin{equation}
S_{F}^{\Lambda} =  {S}_{F1p}^{\Lambda}
+{S}_{F10}^{\Lambda}
+{S}_{F20}^{\Lambda}
+{S}_{F30}^{\Lambda}
+{S}_{mF}^{\Lambda}
+{S}_{\nu_1}^{\Lambda},
\end{equation}
where
\begin{align}
{S}_{F1p}^{\Lambda}= \frac{1}{2g^2}
\sum_{n_2} \sum_{\hat{a}, \hat{b}}
\sum_{{p}_1 = - \Lambda}^{\Lambda} \biggl[&
i\left(
i {p}_1 \omega 
+ i \frac{\alpha^{\hat{a}}_{n_2} -  \alpha^{\hat{b}}_{n_2}}{R_1}
\right)
\hat{\eta}^{\hat{b}\hat{a}}_{+;n_2, -{p}_1} 
\hat{\psi}_{-1;n_2, {p}_1}^{\hat{a}\hat{b}}
\nn \\
&+i\left(
i {p}_1 \omega 
+ i \frac{\alpha^{\hat{a}}_{n_2} -  \alpha^{\hat{b}}_{n_2}}{R_1}
\right)
\hat{\eta}^{\hat{b}\hat{a}}_{-;n_2, -{p}_1} 
\hat{\psi}_{+1;n_2, {p}_1}^{\hat{a}\hat{b}}
\nn \\
&+\left(i {p}_1 \omega + i 
\frac{\alpha^{\hat{a}}_{n_2} -  \alpha^{\hat{b}}_{n_2+1}}{R_1}\right)
\hat{\xi}_{-;n_2, -{p}_1}^{\dag\hat{b}\hat{a}}
\hat{\xi}^{\hat{a}\hat{b}}_{+;n_2, {p}_1}  
\nn \\
&-\left(i {p}_1 \omega + i 
\frac{\alpha^{\hat{a}}_{n_2+1} -  \alpha^{\hat{b}}_{n_2}}{R_1}\right)
\hat{\xi}_{-;n_2, -{p}_1}^{\hat{b}\hat{a}}
\hat{\xi}^{\dag\hat{a}\hat{b}}_{+;n_2, {p}_1}  
\biggr],
\end{align}
\begin{align}
{S}_{F10}^{\Lambda}= \frac{1}{g^2}
\sum_{n_2} \tr  
\biggl[&
+i\left(\hx_{-;n_2} \left(
\hY^\dag_{n_2} \hps_{+1;n_2} -\hps_{+1;n_2+1}\hY^\dag_{n_2} 
\right)\right)_0
-i\left(\hx_{-;n_2}^\dag 
\left(
\hY_{n_2} \hps_{+1;n_2+1}- \hps_{+1;n_2} \hY_{n_2}
\right)\right)_0
\nn \\
&-i\left(\hx_{+;n_2} 
\left(
\hY^\dag_{n_2} \hps_{-1;n_2}- \hps_{-1;n_2+1}\hY^\dag_{n_2}
\right)\right)_0
+i\left(\hx_{+;n_2}^\dag 
\left(
\hY_{n_2} \hps_{-1;n_2+1} -\hps_{-1;n_2} \hY_{n_2}
\right)\right)_0
 \biggr],
\end{align}
\begin{align}
{S}_{F20}^{\Lambda}=
\frac{1}{g^2} \sum_{n_2}
\tr \biggl[ 
&-\frac{1}{2}
\left(
\left(\he_{-;n_2} \hY_{n_2} - \hY_{n_2} \he_{-;n_2+1}\right)
\hx_{+;n_2}^\dag\right)_0
-\frac{1}{2}
\left(
\left(
\he_{-;n_2+1} \hY^\dag_{n_2}
- \hY^\dag_{n_2}\he_{-;n_2}
\right) \hx_{+;n_2}\right)_0
\nn \\
&-\frac{1}{2}\left(
\left(
\he_{+;n_2} \hY_{n_2}-\hY_{n_2} \he_{+;n_2+1} 
\right) \hx_{-;n_2}^\dag\right)_0
-\frac{1}{2}
\left(
\left(
\he_{+;n_2+1} \hY^\dag_{n_2}- \hY^\dag_{n_2}\he_{+;n_2}
\right) \hx_{-;n_2}\right)_0
\nn \\
&
+ \left(\hps_{-1;n_2}[\hC_{n_2}, \hps_{+1;n_2}]
\right)_0
- \left(
\hps_{-1;n_2}[\hph_{+;n_2}, \hps_{-1;n_2}]
\right)_0
+ \left(
\hps_{+1;n_2}[\hph_{-;n_2}, \hps_{+1;n_2}]
\right)_0
\nn \\
&
- \frac{1}{4}
\left(
\he_{+;n_2}[\hC_{n_2}, \he_{-;n_2}]
\right)_0
- \frac{1}{4}\left(
\he_{-;n_2}[\hph_{+;n_2}, \he_{-;n_2}]
\right)_0
+ \frac{1}{4}\left(
\he_{+;n_2}[\hph_{-;n_2}, \he_{+;n_2}]
\right)_0
\biggr],
\end{align}
\begin{align}
{S}_{F30}^{\Lambda}=
\frac{1}{g^2} \sum_{n_2}
\tr \biggl[ 
&
-\frac{1}{2}\left(\hx_{-;n_2}^\dag 
\left(
\hph_{+;n_2}\hx_{-;n_2}-\hx_{-;n_2}\hph_{+;n_2+1}
\right)
\right)_0
-\frac{1}{2}\left(\hx_{-;n_2} 
\left(
\hph_{+;n_2+1} \hx_{-;n_2}^\dag- \hx_{-;n_2}^\dag \hph_{+;n_2}
\right)\right)_0
\nn \\
&+\frac{1}{2}\left(\hx_{+;n_2}^\dag 
\left(
\hph_{-;n_2} \hx_{+;n_2}- \hx_{+;n_2}\hph_{-;n_2+1}
\right)\right)_0
+\frac{1}{2}\left(\hx_{+;n_2} 
\left(
\hph_{-;n_2+1} \hx_{+;n_2}^\dag- \hx_{+;n_2}^\dag\hph_{-;n_2}
\right)\right)_0
\nn \\
&+\frac{1}{2}\left(\hx_{-;n_2}^\dag 
\left(
\hC_{n_2} \hx_{+;n_2}- \hx_{+;n_2}\hC_{n_2+1}
\right)\right)_0
+\frac{1}{2}\left(\hx_{-;n_2} 
\left(
\hC_{n_2+1} \hx_{+;n_2}^\dag- \hx_{+;n_2}^\dag \hC_{n_2}
\right)\right)_0\biggr],
\end{align}
\begin{align}
S_{mF}^{\Lambda} =
\frac{1}{g^2} \sum_{n_2}
\tr \biggl[ 
\frac{2M}{3}\left(\hps_{+1;n_2}\hps_{-1;n_2}\right)_0
+\frac{M}{3}\left(\hx_{+;n_2}\hx_{-;n_2}^\dag\right)_0
+\frac{M}{3}\left(\hx_{+;n_2}^\dag \hx_{-;n_2}\right)_0
-\frac{M}{6}\left(\he_{+;n_2}\he_{-;n_2} \right)_0
\biggr],
\nn \\
\end{align} 
\begin{align}
S_{\nu_1}^{\Lambda} = 
\frac{1}{g^2} \sum_{n_2}
\tr \biggl[ 
&2a^2 \nu_{1}\left(\left( -\hx_{-;n_2} \hx_{+;n_2}^{\dag} 
+ \hx_{+;n_2} \hx_{-;n_2}^{\dag} 
\right)\hat{\cal Y}_{n_2}\right)_0
\nn \\
&-2a^2 \nu_{1}\left(\left( \hx_{-;n_2} \hY_{n_2}^{\dag} 
+ \hY_{n_2} \hx_{-;n_2}^{\dag} 
\right)\left( \hx_{+;n_2} \hY_{n_2}^{\dag} 
+ \hY_{n_2} \hx_{+;n_2}^{\dag} 
\right)
\right)_0
\biggr].
\end{align}
Then we complete the construction of the hybrid regularization theory.
For later discussion, we should keep in mind that
there are no derivative couplings 
other than the 4-point bosonic vertices 
with one derivative 
in $S_{B\nu_1 p}$.

\subsection{Completing the non-perturbative formulation by uplifting the two-dimensional theory to the four-dimensional theory}
\label{Uplift}
Let us complete the construction of the non-perturbative formulation 
for non-commutative ${\cal N} = 2$ four-dimensional $U(k)$ supersymmetric 
Yang-Mills theories.
In the hybrid regularization theory (\ref{Eq:cut-off action}), 
we expand fields around the following
minimum of $k$-coincident Fuzzy 2-sphere,
\begin{equation}
C_{n_2,c} = \frac{2M}{3}L_3, \quad
\phi_{\pm;n_2,c} = \frac{M}{3}\left( L_{1} \pm iL_2 \right), \qquad
\label{Fuzzy S2 solution}
\end{equation}
where $m\times m$ matrices
$L_{\tilde{i}}\, (\tilde{i} = 1,2,3)$ are decomposed to tensor products of $l \times l$ 
and $k \times k$ as
\begin{equation}
L_{\tilde{i}} = L_{\tilde{i}}^{(l)} \otimes {\bf 1}_k, \quad m = lk.
\end{equation}
Here the $L_{\tilde{i}}^{(l)}$ are $SU(2)$ generators in the 
$l (= 2j+1)$-dimensional 
irreducible representation, 
\begin{equation}
[L_{\tilde{i}}^{(l)}, L_{\tilde{j}}^{(l)}] = 
i\epsilon_{\tilde{i}\tilde{j}\tilde{k}} L_{\tilde{k}}^{(l)}.
\end{equation}
The above 
solution uplifts the theory from two dimensions
to four dimensions.\footnote{Good 
references about Fuzzy 2-sphere described here are
\cite{Iso:2001mg,Maldacena:2002rb,Ishii:2008ib},}
The emergent two dimensions by the (\ref{Fuzzy S2 solution})
are regularized by the non-commutative parameter $\Theta$, UV cut-off 
$\hat{\Lambda}$
and radius of the sphere 
$R_f$.
Here the non-commutative parameter is $\Theta = \frac{18}{M^2 l}$ and 
the UV cut-off is
$\hat{\Lambda} \sim 2j/R_f \sim \frac{M}{3}\cdot 2j$, and 
$R_f = \frac{3}{M}$.
The four-dimensional space is consisting of 
$\R^1_{\Lambda}$ which is regularized by the momentum cut-off,  
$\R^1_{orb}$ which is regularized by the orbifold projection, 
and the Fuzzy $S^2$. 
The final form of our non-perturbative formulation 
for the four-dimensional ${\cal N} = 2$ supersymmetric Yang-Mills theories
is obtained by the expansion of fields around the 
(\ref{Fuzzy S2 solution}) in the hybrid regularization theory. 

Let us see how the field variables in the hybrid regularization theory are
expanded around the solution 
(\ref{Fuzzy S2 solution}).
A two-dimensional momentum 
$\hat{\bf p}$ modes on $\R^1_\Lambda \times \R^1_{orb}$ 
of field variables in the $U(m)$ 
hybrid regularization theory, say $B$, 
are expanded further by the spherical harmonics,
\begin{equation}
\tilde{B}(\hat{\bf p}) = \sum_{J = 0}^{2j} \sum_{J_3 = -J}^{J} 
\tilde{h}^{(jj)}_{JJ_3} \otimes b_{JJ_3}(\hat{\bf p}),
\label{Eq:expansion}
\end{equation}
where $\tilde{h}^{(jj)}_{JJ_3}$ is an $l \times l$ matrix corresponding 
to the Fuzzy spherical harmonic, and $b_{JJ_3}(\hat{\bf p})$ is a $k \times k$
matrix becoming a field variable on the target four-dimensional theory.
We truncate the sum of the spherical harmonic over the spin at the level 
spin $j$, the $j$ gives the UV cut-off for 
the Fuzzy 2-sphere directions 
as $\hat{\Lambda} = 2j/R_f$. 

By this uplifting, we have completed the non-perturbative formulation for the 
non-commutative ${\cal N} = 2$ four-dimensional supersymmetric Yang-Mills
theories on $\R^2 \times \R^2_{\Theta}$. In the formulation, 
the four dimensions are regularized as $\R^1_ {\Lambda} \times \R^1_{orb}
\times \text{Fuzzy} \,\, S^2$.

\section{How to take the target theory limit}
In this section, we will explain how we take the target
non-commutative ${\cal N} = 2$ supersymmetric Yang-Mills theory limit
from our non-perturbative formulation constructed in the 
previous section. 
We will explain how the following steps lead to the target theory:
\begin{enumerate}
\item $\Lambda \to \infty$. 
\item $a \to 0$ with $aN$ and $\nu_1$ fixed.
\item $\nu_1 \to 0$,  $aN \to \infty$ .
\item $l, m \to \infty$, ($M \to 0, \hat{\Lambda}, R_f \to \infty$) 
with $k$, $\Theta$ fixed.
\end{enumerate}

\subsection{1st step: From the non-perturbative formulation on 
$\R^1_{\Lambda} \times \R^1_{orb} \times \text{Fuzzy}\,\,S^2$
to the orbifold lattice theory on $\R^1 \times \R^1_{orb} \times \text{Fuzzy}\,\,S^2$}
\label{sec:1d correction}
Here we will explain how the orbifold lattice gauge theory 
(\ref{def:orbifold action})
is recovered from the starting non-perturbative formulation
only by taking the $\Lambda \to \infty$
without any parameter fine-tunings.
Although a one-dimensional system is expected not to have UV divergences
in general, 
there is also a case that dangerous UV divergences requiring fine-tunings 
show up even in one-dimensions as discussed in 
\cite{Giedt:2004vb,Catterall:2000rv}.
So to rigorously confirm the absence of dangerous quantum corrections, 
we will carefully discuss the UV divergences here.

Here we will check the all diagrams with UV divergences,
since all quantum corrections 
come from diagrams with UV divergences. 
In the non-perturbative formulation action, 
which is (\ref{Eq:cut-off action}) expanded around the 
Fuzzy 2-sphere solution (\ref{Fuzzy S2 solution}), 
only the bosonic 4-point vertices in $S_{B\nu_1 p}$ 
can be derivative couplings.
Then the degree of UV divergence $D$ of each diagram
is estimated as follows,
\begin{equation}
D = L_{oop} - 2I_B - I_F + V_{d4}, \qquad 
L_{oop} = 1 + I_B +I_F -V_{d4}-V_{bff}-V_{bbff}-\sum_{n\ge 3}V_{n} ,
\end{equation}
namely
\begin{equation}
D = 1 -I_B-V_{bff}-V_{bbff}-\sum_{n\ge 3}V_{n} .
\end{equation}
And 
\begin{equation}
E_B + 2I_B = \sum_{n \ge 3} (n V_{n} )
+4V_{d4}+ V_{bff}+2V_{bbff}, \qquad  E_F + 2I_F = 2 V_{bff}+2V_{bbff} .
\end{equation}
Here $L_{oop}$ is the number of loops and each $I_B$ and $I_F$ is 
the number of internal lines of bosons and fermions respectively.
The $V_{n} (n\ge 3)$ are the number of bosonic $n$-point vertices 
without derivative and $V_{d4}$ is the number of bosonic 
4-point vertices with one derivative, 
each $V_{bff}$ and $V_{bbff}$ is the number 
of boson-fermion-fermion vertices and 
boson-boson-fermion-fermion interaction terms respectively.
And each $E_B$ and $E_F$ is the number of external 
bosons and fermions respectively.
Only following three diagrams can have UV divergences with $D \ge 0$,
\begin{equation}
I_F = E_B = V_{bff} = 1,  \quad E_F = I_B = V_{n} = V_{d4} = V_{bbff}= 0, 
\label{Fermion loop 1}
\end{equation}
\begin{equation}
E_B = 2, \quad I_F =  V_{bbff} = 1,  \quad 
E_F = I_B = V_{n} = V_{d4} = V_{bff} = 0, 
\label{Fermion loop 2}
\end{equation}
and 
\begin{equation}
E_B =2, \quad V_{d4} = I_B = 1, \quad 
V_{bff} =V_{bbff} =  E_F = V_{n} = I_F = 0.
\label{bosonic loop}
\end{equation}
(\ref{Fermion loop 1}) 
is the bosonic tadpole diagram with fermionic 1-loop,
and (\ref{Fermion loop 2}) is bosonic 2-point function with fermionic 1-loop
whose vertex is a boson-boson-fermion-fermion vertex.
And (\ref{bosonic loop}) is the bosonic 2-point function with bosonic 
4-point derivative coupling and bosonic 1-loop. 
The UV divergent parts 
of the (\ref{Fermion loop 1}) and (\ref{Fermion loop 2}) 
are 
\begin{equation}
\sim \int^{\Lambda}_{-\Lambda} dk_1 \frac{1}{k_1}
\end{equation}
where $k_1$ is one-dimensional momentum and 
the $\frac{1}{k_1}$ comes from the fermion propagator.
And the UV divergent part of (\ref{bosonic loop}) is
\begin{equation}
\sim \int^{\Lambda}_{-\Lambda} dk_1 \frac{k_1}{k_1^2}
\end{equation}
where the factor in the denominator $k_1^2$ 
comes from the bosonic propagator and 
the factor $k_1$ in the numerator comes from the derivative coupling.
So we can see that all the UV divergent parts of these diagrams are the 
momentum integration of the odd function of the momentum.
Then if we set the integration domain as symmetric,
$[-\Lambda, \Lambda]$, the UV divergent parts
become zero. 
They will not give even finite corrections at 
$\Lambda \to \infty$.
So we can see 
that there are no quantum corrections blocking from reaching the 
orbifold lattice gauge theory 
(\ref{def:orbifold action})
at $\Lambda \to \infty$.

In the appendix~\ref{soft breaking case},
we also discuss the $\Lambda \to \infty$ limit in the case  
of employing the soft SUSY breaking mass terms (\ref{soft breaking mass term}).

Then we have shown 
that the orbifold lattice theory on $\R^1 \times \R^1_{orb} \times \text{Fuzzy}\,\, S^2$ is obtained from the starting non-perturbative formulation on 
$\R^1_{\Lambda} \times \R^1_{orb} \times \text{Fuzzy}\,\, S^2$ by the 1st step
without any fine-tunings.

\subsection{2nd and 3rd steps: From the orbifold lattice theory to 
the non-commutative supersymmetric Yang-Mills on $\R^2 \times \text{Fuzzy}\,\,S^2$.}
\label{Sec:orb-continuum}
Next let us discuss the 2nd and 3rd steps starting from 
the orbifold lattice theory.
The tree level target continuum theory of the orbifold theory
(\ref{def:orbifold action}) is 
\begin{align}
S_{2d,\nu_1} = \frac{2}{g_{2d}^2}\int d^2 x \,   \tr \biggl[ 
&\frac{1}{2}F_{12}^2
+\frac{1}{2}D_{\mu}s^iD_{\mu}s^i
-\frac{1}{4}[s^i, s^j]^2 
+\frac{i}{2}\Psi^T \gamma^{i} [s_i, \Psi]
+\frac{1}{2}\Psi^T (D_1 + \gamma_2 D_2) \Psi
\nn \\
&
-i\frac{M}{6}{\Psi}^T\gamma^{23}{\Psi} 
+\left(\frac{M}{3} + 4 \nu_1
\right)i{s}^3F_{12} +\frac12\left(\frac{M}{3}\right)^2 ({s}^a)^2
 +i\frac{M}{3}\epsilon_{abc}{s}^a{s}^b{s}^c
\nn \\
&+\left( -8 \nu_1^2 - \frac{2M\nu_1}{3}\right) 
s_3^2
-4\nu_1 \chi_{-} \chi_{+}
\biggr].
\label{Target continuum}
\end{align}
(Here we described $SU(m)$ part only.)
Let us confirm possible perturbative quantum
corrections, which could prevent from reaching the target continuum theory, are absent. 
Since the moduli fixing terms (\ref{SUSY preserving mass term}) 
do not break 
$Q_{\pm}$ and the $SU(2)_R$ symmetry,
the argument goes completely parallel to the one in 
\cite{Kaplan:2002wv,Hanada:2011qx}.
We will consider local operators with positive mass dimension $p$ 
near the continuum limit,
\begin{equation}
{\cal O}_p = \tilde{M}^q \varphi^\alpha \del^\beta \psi^{2 \gamma}, 
\qquad p = q + \alpha + \beta + 3 \gamma ,
\end{equation}
where $\varphi$, $\psi$ and $\del$ denote 
bosonic fields, fermions fields, and derivatives, respectively.
$\tilde{M}$ represents mass parameter $M$ or $\nu_1$.
And $q,\alpha,\beta,\gamma = 0,1,2, \cdots$.

From dimensional analysis,
radiative corrections 
to the ${\cal O}_p$ have the form
\begin{equation}
\left( 
\frac{1}{g_{2d}^2}c_0 a^{p-4} + c_1 a^{p-2} + g_{2d}^2 a^p + \ldots 
\right)\int d^2 x {\cal O}_p \label{counter term}
\end{equation}
where the first, second and third terms and $"\ldots"$ in the parenthesis are 
contributions from tree level, 1-loop, 2-loop and higher-loops
respectively.
The $c_0, c_1, c_2$ are the dimensionless numerical constants.
Since relevant or marginal operators 
appear with the negative power of lattice spacing, we only have to 
focus on terms up to 1-loop order. 
In order for 1-loop coefficients $c_1 a^{p-2}$
to be relevant or marginal, $p$ must be $p =1,2$.
Only the operators $\tr (\varphi), \tilde{M} \varphi, \varphi^2$ 
can be mass dimensions $p =1,2$ among the dynamical operators.
But these candidates 
are not allowed to show up
due to the preserved supersymmetry $Q_{\pm}$ and $SU(2)_R$
on the lattice.
So, in the 2nd step $a \to 0$, 
the quantum continuum theory (\ref{Target continuum}) is obtained 
without any fine-tuning.

Because there are no dynamical relevant or marginal operators which respect 
neither 8 supersymmetry nor $SO(2)$ rotational symmetry, 
the both symmetries are 
recovered automatically only by taking 3rd step $\nu_1 \to 0$ 
in the quantum continuum theory 
(\ref{Target continuum}).
By taking the 3rd step $\nu_1 \to 0$, we obtain 
the non-commutative 
supersymmetric Yang-Mills
theory on $\R^2 \times \text{Fuzzy}\,\,S^2$,
\begin{align}
S_{2d,0} = \frac{2}{g_{2d}^2}\int d^2 x \, \tr \biggl[& 
\frac{1}{2}F_{12}^2
+\frac{1}{2}D_{\mu}s^iD_{\mu}s^i
-\frac{1}{4}[s^i, s^j]^2 
+\frac{i}{2}\Psi^T \gamma^{i} [s_i, \Psi]
+\frac{1}{2}\Psi^T (D_1 + \gamma_2 D_2) \Psi
\nn \\
&
-i\frac{M}{6}{\Psi}^T\gamma^{23}{\Psi} 
+\frac{iM}{3}{s}^3F_{12} +\frac12\left(\frac{M}{3}\right)^2 ({s}^a)^2
 +i\frac{M}{3}\epsilon_{abc}{s}^a{s}^b{s}^c
\biggr],\nn \\ 
\label{Eq: 2d continuum theory 2}
\end{align}
with full 8 SUSY and the $SO(2)$ rotational symmetry on $\R^2$.

Although the orbifold lattice theory on $\R^1 \times \R^1_{orb}
\times \text{Fuzzy}\,\, S^2$ 
is an anisotropic one which does not possess
even finite point subgroups of $SO(2)$ rotational 
symmetry of (\ref{Eq: 2d continuum theory 2}), 
possible 
$SO(2)$ breaking corrections are all irrelevant thanks 
to the super-renormalizability and other preserved symmetries 
on the lattice~\cite{Kaplan:2002wv}.

Even if we introduce the soft SUSY breaking moduli fixing terms
(\ref{soft breaking mass term}) instead of the SUSY preserving terms
(\ref{SUSY preserving mass term}),
discussion will not be changed.
Because they are just soft breaking terms
which will not alter UV divergences.

Then we have shown that the 2nd and 3rd steps 
also do not require any fine-tunings.

\subsection{Final step: From the theory on Fuzzy 2-sphere to the one on 
the Moyal plane}
\label{Sec:fourth-step}

After the first three steps, 
we will undertake to manage the Fuzzy sphere 
directions.
Here we will take the 
$\hat{\Lambda} \to \infty$ by taking $l \to \infty$
with fixed $\Theta$ and $k$.
In the limit, the Fuzzy sphere is decompactified 
to be flat Moyal plane $\R_{\Theta}^2$ because 
\begin{equation}
R_F \sim 1/M \sim l^{1/2} \to \infty, \qquad
\hat{\Lambda} \sim l^{1/2} \to \infty. \label{Eq: Fuzzy to com. limit}
\end{equation}
By taking the limit,
the theory becomes the ${\cal N} = 2$ $U(k)$ four-dimensional
gauge theory on $\R^2 \times \R^2_{\Theta}$.
The four-dimensional gauge coupling is given by the non-commutative 
parameter as
\begin{equation}
g_{4d}^2 = 2 \pi \Theta g_{2d}^2.
\end{equation}
After taking the limit (\ref{Eq: Fuzzy to com. limit}),
the Fuzzy spherical harmonic expansion (\ref{Eq:expansion})
can be transcribed to the 
one by the plane waves on $\R_{\Theta}^2$ as
\begin{equation}
\tilde{B}({\bf p}) = 
\int \frac{d^2 \tilde{p}}{(2\pi)^2} e^{i \tilde{p}\cdot \hat{x}} \otimes 
\tilde{b}({\bf p}, \tilde{p}).
\end{equation}
The set 
$({\bf p}, \tilde{p})$ represents the four-momenta 
on $\R^2 \times \R^2_{\Theta}$
where ${\bf p}$ is the two-momenta on $\R^2$ coming from $\hat{\bf p}$ 
on $\R^1_{\Lambda} \times \R^1_{orb}$, 
and $\tilde{p}$ is the two-momenta
on $\R^2_{\Theta}$.
$\tilde{h}^{(jj)}_{JJ_3}$ in (\ref{Eq:expansion}) 
are transcribed into the plane waves 
$e^{i\tilde{p} \cdot \hat{x}}$ on $\R^2_{\Theta}$, and the modes
$\tilde{b}({\bf p}, \tilde{p})$ on $\R^2 \times \R^2_{\Theta}$
come from the $k \times k$ matrix $b_{JJ_3}({\bf p})$.
The inner product $\widetilde{Tr}$ 
between the plane waves on $\R^2_{\Theta}$ is
defined as
\begin{equation}
\widetilde{Tr}\left( 
e^{i \tilde{p}\cdot \hat{x}}
e^{i \tilde{q}\cdot \hat{x}} \right) = \frac{2\pi }{\Theta}
\delta^2(\tilde{p}+\tilde{q}),
\end{equation}
where the inner product depends on the $\Theta$.

Then finally we have shown that the target
non-commutative four-dimensional ${\cal N} = 2$ supersymmetric theory
on $\R^2 \times \R^2_{\Theta}$
can be reached with no fine-tunings 
by taking the following steps:
\begin{enumerate}
\item $\Lambda \to \infty$. 
\item $a \to 0$ with $aN$ and $\nu_1$ fixed.
\item $\nu_1 \to 0$,  $aN \to \infty$ .
\item $l, m \to \infty$, ($M \to 0, \hat{\Lambda}, R_f \to \infty$) 
with $k$, $\Theta$ fixed.
\end{enumerate}

If $\Theta \to 0$ can be continuously connected 
to the commutative theory,
our formulation can be a non-perturbative formulation also for the 
commutative
${\cal N} = 2$ supersymmetric Yang-Mills theories.
But we should note that more investigation is needed to clarify 
whether $\Theta \to 0$ can be continuously connected to the commutative
theories or not.


\section{Summary}
In this paper we proposed a non-perturbative formulation for the 
non-commutative 
${\cal N} =2$ four-dimensional supersymmetric Yang-Mills theories.
We made the formulation from the one-dimensional mass-deformed 
matrix model with 8 supercharges by performing the regularization 
which is a combination of the orbifold projection, the momentum 
cut-off and the generation of the Fuzzy 2-sphere. 
Similar to the formulation in \cite{Hanada:2011qx},
our formulation enables numerical studies of
non-commutative four-dimensional ${\cal N} = 2$ supersymmetric Yang-Mills theories on $\R^2 \times \R^2_{\Theta}$. 
The absence of any parameter fine-tuning was confirmed at all order of 
perturbation. 
We should note that non-commutative gauge theory 
plays an important role to clarify non-perturbative aspects of the 
gauge theories. 
Therefore 
our formulation will play a very important role 
to uncover 
non-perturbative structures of the supersymmetric gauge 
theories through numerical studies of ${\cal N} = 2$ non-commutative
supersymmetric Yang-Mills theories.

Our formulation has several advantages.
First this formulation is simpler than the similar model \cite{Hanada:2011qx},
and easier to put on a computer.
Hence it is easy to check the absence of fine-tunings at non-perturbative
level.
Moreover, after we take the $\Lambda \to \infty$ limit,
our formulation can possesses more supersymmetry than
the formulation in \cite{Hanada:2011qx} in the UV region, 
so it would be
easier to control UV divergences.
In contrast to conventional moduli fixing terms in the orbifold lattice 
gauge theories, we proposed a new moduli fixing terms 
(\ref{SUSY preserving mass term})
with keeping a 
partial SUSY on the lattice.
The new fixing terms could replace SUSY breaking mass terms,
and the new terms would make conclusions of simulations more concrete.
These advantages are owed to the highly anisotropic nature of the 
formulation. This demonstrates that anisotropic 
treatments can be very efficient
to deal with four-dimensional theories.

The $\Theta \to 0$ limit of the four-dimensional ${\cal N} =2$ 
supersymmetric 
Yang-Mills theory is expected not to be continuously connected to the
commutative ${\cal N} =2$ theory due to the 
UV/IR mixing~\cite{Minwalla:1999px}.
This is in 
contrast to the ${\cal N} = 4$ four-dimensional supersymmetric Yang-Mills
theories. 
There is a discussion, however, 
that the non-commutative four-dimensional ${\cal N}=2$ theory may flow to 
the ordinary commutative theory in the infrared \cite{Armoni:2001br}. 
So there is still a chance for our formulation to be a numerical 
tool also for the commutative four-dimensional ${\cal N}=2$ theories.
To investigate the possibility, 
it would be also interesting to do numerical studies of behavior
of infrared quantities on the $\R^2 \times \R^2_{\Theta}$.
These studies might be able to provide some insight to reach
the commutative ${\cal N} =2$ supersymmetric Yang-Mills theories.

The two-dimensional mass-deformed theory with 8 supercharges 
itself is also interesting on its own.
In the previous simulations, and
in particular simulations on two-dimensional deconstruction formulations, 
the conclusion 
has been more or less obscured
because SUSY is broken by the moduli fixing terms. 
But by the new moduli fixing terms (\ref{SUSY preserving mass term}), 
it would be possible to get more concrete result 
since the new terms keep the lattice supersymmetry.

It is also interesting to study about the relationship between 
the mechanism for uplifting flat-directions
and the $\Omega$-deformation 
\cite{Nekrasov:2003af, Nekrasov:2003rj}.
The $\Omega$-deformation has been 
used to regularize the instanton moduli space of four-dimensional
${\cal N} =2$ supersymmetric Yang-Mills theory, 
and the deformation uplifts the flat moduli space with keeping the SUSY.
In fact, in \cite{Ohta:2006qz}, it has been shown 
that a deconstruction lattice 
formulation for the two-dimensional ${\cal N} = (4,4)$ supersymmetric 
gauge theory 
can be embedded into the topological matrix model with $\Omega$-deformation 
in the \cite{Nekrasov:2002qd},
and the flat directions of the deconstruction theory are uplifted
by the $\Omega$-background with keeping the SUSY. 
Studies on the relationship of lattice formulations 
to such a mathematical method would provide 
a clue as to how 
to construct more sophisticated formulations for numerical studies
of supersymmetric gauge theories.

\section*{Acknowledgment}
The author is deeply grateful to 
M.~Hanada and S.~Matsuura 
for valuable discussions and for giving many useful comments.
He also would like to thank M.~Abbott and D.~Elander for reading the 
manuscript very carefully and giving useful comments.
T.T would like to thank ICTP, Isaac Newton Institute for Mathematical Sciences,
NCTS, and National Taiwan University for hospitality during his stay.

\appendix
\section{Gamma matrices}
\label{Sec:gamma matrix}
The gamma matrices 
$\gamma_i$ ($i=2,\cdots,6$) are
$8\times 8$ matrices satisfying $\{\gamma_i,\gamma_j\}=-2\delta_{ij}$ 
and $\gamma_2\cdots\gamma_6=-i {\bf 1}_8$.  
Their explicit form we use is 
\begin{align}
 \gamma_2 &= -i\left(\begin{matrix}
\sigma_3&&&\\
&\sigma_3&&\\
&&\sigma_3&\\
&&&\sigma_3\end{matrix}\right), \quad 
\gamma_3=\left(\begin{matrix}
&-\sigma_2&& \\
\sigma_2&&& \\
&&&-\sigma_2 \\
&&\sigma_2&
\end{matrix}\right), \nn \\ 
\gamma_4&=-i\left(\begin{matrix}
\sigma_1&&& \\
&\sigma_1&& \\
&&\sigma_1& \\
&&&\sigma_1
\end{matrix}\right), \quad 
\gamma_5=\left(\begin{matrix}
&&-\sigma_2& \\
&&&\sigma_2 \\\
\sigma_2&&& \\
&-\sigma_2&&
\end{matrix}\right), \quad 
\gamma_6=\left(\begin{matrix}
&&&-\sigma_2 \\
&&-\sigma_2& \\
&\sigma_2&& \\
\sigma_2&&&
\end{matrix}\right), 
\label{gamma}
\end{align}
where $\sigma_{1,2,3}$ are Pauli matrices.

\section{Gauge fixing}
\label{Sec:gauge fixing}
The orbifold lattice gauge theory is an $mN \times mN$ 
matrix model with $U(m)^N$ gauge symmetry.
The $N$ gauge fields $v^1_{n_2}(x_1)$ as well as 
$U(m)^N$ gauge transformation matrix $U_{n_2}$ are 
embedded in an $mN \times mN$ matrix block diagonally as
\begin{equation}
v^1(x_1) 
=
\begin{pmatrix}
v^1_{1}(x_1) & &   \\
& v^1_{2}(x_1) &    \\
& & \ddots &    \\
& & & v^1_N(x_1)   
\end{pmatrix},
\qquad
U(x_1) 
=
\begin{pmatrix}
U_{1}(x_1) & &   \\
& U_{2}(x_1) &    \\
& & \ddots &    \\
& & & U_N(x_1)   
\end{pmatrix}.
\label{Block diagonally}
\end{equation}
All site fields are embedded in the above manner.
Here we will consider following loop operators $W^c_{n_2}$ 
on a site $n_2$,
\begin{equation}
W^{c}_{n_2} (x_1) \equiv  
\exp \left(i \int_{x_1}^{x_1 + R_1} dx_1' v^1_{n_2}(x_1') \right),
\end{equation}
which transform under the $U(m)^N$ in the adjoint representation,
\begin{equation}
U_{n_2}(x_1)W^c_{n_2}(x_1)U_{n_2}(x_1)^\dag.
\end{equation}
The loop operators are also embedded into an $mN \times mN$ matrix
block diagonally in the same way as (\ref{Block diagonally}).
Here the $x_1$ direction is taken to be periodic,
$x_1 \sim x_1 + R_1$.
Next we define the $N$ sets of $m \times m$ 
constant diagonal matrices $\bar{\alpha}_{n_2}$ residing on the sites,
\begin{equation}
\bar{\alpha}_{n_2} 
=
\diag (e^{i\alpha_{1n_2}}, \ldots,e^{i\alpha_{\hat{a}n_2}}, \ldots 
e^{i\alpha_{mn_2}}).
\end{equation}
These $\bar{\alpha}_{n_2}$ are constant with respect to the $x_1$, while
they are not constant with respect to $n_2$.
And they should satisfy 
$\sum_{n_2} \sum_{\hat{a}} \alpha_{n_2,\hat{a}} = 0$ to make them 
consistent with the gauge fields which is traceless in the sense of 
$mN \times mN$ matrix $\Tr (v^1) = \sum_{n_2} \tr (v^1_{n_2}) = 0$.
By using $W^c$ and $\bar{\alpha}$, we will add the 
gauge fixing terms in the BRS exact form
\begin{align}
&\frac{1}{g^2}{\cal Q} \int dx_1 \sum_{n_2} \tr 
\left( 
\bar{c}_{n_2}(x_1) \left( 
W^c_{n_2}(x_1) -\bar{\alpha}_{n_2} 
\right)
\right)
\nn \\
=&
\frac{1}{g^2}\int dx_1 \sum_{n_2} \tr 
\left( 
{\cal B}_{n_2}(x_1) \left( 
W^c_{n_2}(x_1) -\bar{\alpha}_{n_2}
\right)
\right)
+\frac{i}{g^2}\int dx_1 \sum_{n_2} \tr 
\left( \bar{c}_{n_2}(x_1)
[\bar{\alpha}_{n_2}, c_{n_2}(x_1)]
\right),
\label{gauge fixing term}
\end{align}
where ${\cal Q}$ is the BRS charge different from the supercharges.
Here all ghost $c_{n_2}$, 
anti-ghost $\bar{c}_{n_2}$ and Nakanishi-Lautrup fields ${\cal B}_{n_2}$ 
are sitting on the sites.
If we explicitly write down the FP term
with the $m\times m$ indices, it becomes
\begin{align}
&\frac{i}{g^2}\int dx_1 \sum_{n_2} \sum_{\hat{a} \ne \hat{b}}
\left(
\bar{c}^{\hat{b}\hat{a}}_{n_2}(x_1) 
(e^{i\alpha_{\hat{a}n_2}}- e^{i\alpha_{\hat{b}n_2}} )
c^{\hat{a}\hat{b}}_{n_2}(x_1) \right)
\nn \\
=& -\frac{2}{g^2}\int dx_1 \sum_{n_2} \sum_{\hat{a} \ne \hat{b}}
\left[\exp \left(\frac{i\alpha_{\hat{a}n_2}+i\alpha_{\hat{b}n_2}}{2}\right) 
\sin \left(\frac{\alpha_{\hat{a}n_2}-\alpha_{\hat{b}n_2}}{2}\right)
\bar{c}^{\hat{b}\hat{a}}_{n_2}(x_1)c^{\hat{a}\hat{b}}_{n_2}(x_1)\right].
\end{align}
Here due to the property 
$\sum_{n_2} \sum_{\hat{a}} \alpha_{n_2,\hat{a}} = 0$,
it becomes
\begin{equation}
\exp\left(\sum_{n_2}\sum_{\hat{a}\ne \hat{b}} \left( 
\frac{i\alpha_{\hat{a}n_2}+i\alpha_{\hat{b}n_2}}{2}\right) \right) = 1.
\end{equation}
So each momentum mode of the ghost provides the FP determinant term 
(we factor out the parts independent of $\alpha_{\hat{a}n_2}$)
\begin{equation}
\exp \left( - S_{FP}\right) =
\exp\left(2 \sum_{n_2}\sum_{\hat{a} < \hat{b}} \log \left|
\sin \frac{\alpha_{\hat{a}n_2} - \alpha_{\hat{b}n_2}}{2}
\right|\right) .
\end{equation}

Actually the above gauge fixing (\ref{gauge fixing term}) is not
enough to justify the momentum cut-off regularization.
Following $U(1)^{mN}$ transformations of the Cartan elements of the 
$v^1_{n_2}$ remain as 
residual gauge symmetry 
\begin{equation}
v^1_{n_2, \hat{a}\hat{a}} \to 
v^1_{n_2, \hat{a}\hat{a}} + \del_1 \phi_{n_2 \hat{a}} ,
\end{equation}
with 
\begin{equation}
\phi_{n_2 \hat{a}}(x_1 + R_1) - 
\phi_{n_2 \hat{a}}(x_1 )  = 2\pi n_{n_2 \hat{a}},
\end{equation}
where $n_{n_2 \hat{a}} \in \mathbb{Z}$.
These are the 
large gauge transformations generating non-zero winding numbers.
These gauge transformations have an effect
to shift
$v^1_{n_2\hat{a}} \to v^1_{n_2\hat{a}} + 2\pi n_{n_2\hat{a}}$, 
$\alpha_{n_2}^{\hat{a}} \to \alpha_{n_2}^{\hat{a}} + 2\pi n_{n_2\hat{a}}$.
Please note that 
gauge fixing function is unchanged under the shift because of
$e^{i 2\pi n_{n_2\hat{a}}} =1 $.
Through the covariant derivative $D_{1}$ (or ${\cal D}_1$)
this shift is transcribed into the 
shift of the momentum along the $x_1$.
For instance, if we consider the covariant derivative of $Y$,
\begin{equation}
{\cal D}_1 Y_{n_2}^{\hat{a}\hat{b}} \sim
\left(ip_1 \omega 
+ i\frac{\alpha^{\hat{a}}_{n_2} - \alpha^{\hat{b}}_{n_2+1} }{R_1} 
\right)\hat{Y}_{n_2, p_1}^{\hat{a}, \hat{b}},
\end{equation}
after a large gauge transformation, it is transformed as
\begin{align}
\left(ip_1 \omega 
+ i\frac{\alpha^{\hat{a}}_{n_2} - \alpha^{\hat{b}}_{n_2+1} }{R_1} 
\right)
&\to \left(ip_1 \omega 
+ i\frac{(\alpha^{\hat{a}}_{n_2} + 2\pi n_{n_2\hat{a}}) 
- (\alpha^{\hat{b}}_{n_2+1} + 2\pi n_{n_2+1\hat{b}})}{R_1} 
\right)
\nn \\
&=\left(i(p_1 +n_{n_2\hat{a}}-n_{n_2+1\hat{b}})  \omega 
+ i\frac{\alpha^{\hat{a}}_{n_2} 
- \alpha^{\hat{b}}_{n_2+1}}{R_1} 
\right),
\end{align}
then eventually the momentum is shifted as
$p_1 \to p_1 + n_{n_2\hat{a}}-n_{n_2+1\hat{b}}$.
This transformation may spoil the momentum cut-off regularization,
because it allows the momentum to go beyond the momentum cut-off $\Lambda$.
Therefore to justify the momentum cut-off regularization, we need to 
fix these large gauge transformations.
To fix the large gauge transformation, we restrict the domain of the 
$\alpha_{n_2}^{\hat{a}}$ as
\begin{equation}
\max (\alpha_{n_2}^{\hat{a}}) - \min (\alpha_{n_2}^{\hat{b}}) \le 2 \pi  .
\end{equation}

\section{$\Lambda \to \infty$ limit in the case  
of employing the soft SUSY breaking mass terms (\ref{soft breaking mass term})}
\label{soft breaking case}
Let us consider $\Lambda \to 0$ limit 
in the case of employing the 
soft SUSY breaking mass term (\ref{soft breaking mass term})
instead of SUSY preserving terms (\ref{SUSY preserving mass term}).
In this case, there are no derivative couplings
because there are not $S_{B\nu_1 p}$. 
The absence of the derivative couplings are owed to the gauge fixing also.
In this case, the degree of UV divergence of each diagram is estimated as
\begin{equation}
D = L_{oop} - 2I_B - I_F, \qquad L_{oop} = 1 + I_B +I_F -V_{bff}
-\sum_{n} V_{n} ,
\end{equation}
\begin{equation}
E_B + 2I_B = \sum_n \left(nV_{n}\right) 
+ V_{bff}, \qquad  E_F + 2I_F = 2 V_{bff} .
\end{equation}
Only the following one diagram can be $D \ge 0$ with UV divergences,
\begin{equation}
I_F = E_B = V_{bff} = 1,  \qquad E_F = I_B = V_{n}  = 0. 
\label{soft mass divergence}
\end{equation}
The diagram (\ref{soft mass divergence}) is the same as the diagram 
(\ref{Fermion loop 1}),
and its UV divergence is guaranteed to vanish.
So also in this case,
there are no quantum corrections blocking from recovering the 
orbifold lattice gauge theory at $\Lambda \to \infty$.




\begin{thebibliography}{10}

\bibitem{Hanada:2011qx}
M.~Hanada, S.~Matsuura, and F.~Sugino, {\it {Non-perturbative construction of
  2D and 4D supersymmetric Yang-Mills theories with 8 supercharges}},  {\em
  Nucl.Phys.} {\bf B857} (2012) 335--361,
  [\href{http://xxx.lanl.gov/abs/1109.6807}{{\tt 1109.6807}}].

\bibitem{Maru:1997kh}
N.~Maru and J.~Nishimura, {\it Lattice formulation of supersymmetric yang-mills
  theories without fine-tuning},  {\em Int. J. Mod. Phys.} {\bf A13} (1998)
  2841--2856, [\href{http://xxx.lanl.gov/abs/hep-th/9705152}{{\tt
  hep-th/9705152}}].

\bibitem{Giedt:2009yd}
J.~Giedt, {\it {Progress in four-dimensional lattice supersymmetry}},  {\em
  Int. J. Mod. Phys.} {\bf A24} (2009) 4045--4095,
  [\href{http://xxx.lanl.gov/abs/0903.2443}{{\tt 0903.2443}}].

\bibitem{Hanada:2007ti}
M.~Hanada, J.~Nishimura, and S.~Takeuchi, {\it {Non-lattice simulation for
  supersymmetric gauge theories in one dimension}},  {\em Phys.Rev.Lett.} {\bf
  99} (2007) 161602, [\href{http://xxx.lanl.gov/abs/0706.1647}{{\tt
  0706.1647}}].

\bibitem{Anagnostopoulos:2007fw}
K.~N. Anagnostopoulos, M.~Hanada, J.~Nishimura, and S.~Takeuchi, {\it {Monte
  Carlo studies of supersymmetric matrix quantum mechanics with sixteen
  supercharges at finite temperature}},  {\em Phys.Rev.Lett.} {\bf 100} (2008)
  021601, [\href{http://xxx.lanl.gov/abs/0707.4454}{{\tt 0707.4454}}].

\bibitem{Hanada:2008gy}
M.~Hanada, A.~Miwa, J.~Nishimura, and S.~Takeuchi, {\it {Schwarzschild radius
  from Monte Carlo calculation of the Wilson loop in supersymmetric matrix
  quantum mechanics}},  {\em Phys.Rev.Lett.} {\bf 102} (2009) 181602,
  [\href{http://xxx.lanl.gov/abs/0811.2081}{{\tt 0811.2081}}].

\bibitem{Hanada:2008ez}
M.~Hanada, Y.~Hyakutake, J.~Nishimura, and S.~Takeuchi, {\it {Higher derivative
  corrections to black hole thermodynamics from supersymmetric matrix quantum
  mechanics}},  {\em Phys.Rev.Lett.} {\bf 102} (2009) 191602,
  [\href{http://xxx.lanl.gov/abs/0811.3102}{{\tt 0811.3102}}].

\bibitem{Hanada:2009ne}
M.~Hanada, J.~Nishimura, Y.~Sekino, and T.~Yoneya, {\it {Monte Carlo studies of
  Matrix theory correlation functions}},  {\em Phys.Rev.Lett.} {\bf 104} (2010)
  151601, [\href{http://xxx.lanl.gov/abs/0911.1623}{{\tt 0911.1623}}]. 4 pages,
  3 figures.

\bibitem{Hanada:2011fq}
M.~Hanada, J.~Nishimura, Y.~Sekino, and T.~Yoneya, {\it {Direct test of the
  gauge-gravity correspondence for Matrix theory correlation functions}},  {\em
  JHEP} {\bf 1112} (2011) 020.

\bibitem{Berenstein:2008jn}
D.~Berenstein, R.~Cotta, and R.~Leonardi, {\it {Numerical tests of AdS/CFT at
  strong coupling}},  {\em Phys.Rev.} {\bf D78} (2008) 025008,
  [\href{http://xxx.lanl.gov/abs/0801.2739}{{\tt 0801.2739}}].

\bibitem{Berenstein:2010xw}
D.~Berenstein, {\it {Lessons in quantum gravity from quantum field theory}},
  {\em AIP Conf.Proc.} {\bf 1318} (2010) 26--37,
  [\href{http://xxx.lanl.gov/abs/1010.3270}{{\tt 1010.3270}}].

\bibitem{Catterall:2008yz}
S.~Catterall and T.~Wiseman, {\it {Black hole thermodynamics from simulations
  of lattice Yang-Mills theory}},  {\em Phys.Rev.} {\bf D78} (2008) 041502,
  [\href{http://xxx.lanl.gov/abs/0803.4273}{{\tt 0803.4273}}].

\bibitem{Catterall:2009xn}
S.~Catterall and T.~Wiseman, {\it {Extracting black hole physics from the
  lattice}},  {\em JHEP} {\bf 04} (2010) 077,
  [\href{http://xxx.lanl.gov/abs/0909.4947}{{\tt 0909.4947}}].

\bibitem{Catterall:2009it}
S.~Catterall, D.~B. Kaplan, and M.~Unsal, {\it {Exact lattice supersymmetry}},
  {\em Phys.Rept.} {\bf 484} (2009) 71--130,
  [\href{http://xxx.lanl.gov/abs/0903.4881}{{\tt 0903.4881}}]. Invited review
  for Physics Reports. 126 pages.

\bibitem{Cohen:2003xe}
A.~G. Cohen, D.~B. Kaplan, E.~Katz, and M.~Unsal, {\it Supersymmetry on a
  euclidean spacetime lattice. i: A target theory with four supercharges},
  {\em JHEP} {\bf 08} (2003) 024,
  [\href{http://xxx.lanl.gov/abs/hep-lat/0302017}{{\tt hep-lat/0302017}}].

\bibitem{Cohen:2003qw}
A.~G. Cohen, D.~B. Kaplan, E.~Katz, and M.~Unsal, {\it Supersymmetry on a
  euclidean spacetime lattice. ii: Target theories with eight supercharges},
  {\em JHEP} {\bf 12} (2003) 031,
  [\href{http://xxx.lanl.gov/abs/hep-lat/0307012}{{\tt hep-lat/0307012}}].

\bibitem{Kaplan:2005ta}
D.~B. Kaplan and M.~Unsal, {\it A euclidean lattice construction of
  supersymmetric yang- mills theories with sixteen supercharges},  {\em JHEP}
  {\bf 09} (2005) 042, [\href{http://xxx.lanl.gov/abs/hep-lat/0503039}{{\tt
  hep-lat/0503039}}].

\bibitem{Sugino:2003yb}
F.~Sugino, {\it A lattice formulation of super yang-mills theories with exact
  supersymmetry},  {\em JHEP} {\bf 01} (2004) 015,
  [\href{http://xxx.lanl.gov/abs/hep-lat/0311021}{{\tt hep-lat/0311021}}].

\bibitem{Sugino:2004qd}
F.~Sugino, {\it Super yang-mills theories on the two-dimensional lattice with
  exact supersymmetry},  {\em JHEP} {\bf 03} (2004) 067,
  [\href{http://xxx.lanl.gov/abs/hep-lat/0401017}{{\tt hep-lat/0401017}}].

\bibitem{Sugino:2004uv}
F.~Sugino, {\it Various super yang-mills theories with exact supersymmetry on
  the lattice},  {\em JHEP} {\bf 01} (2005) 016,
  [\href{http://xxx.lanl.gov/abs/hep-lat/0410035}{{\tt hep-lat/0410035}}].

\bibitem{Sugino:2006uf}
F.~Sugino, {\it Two-dimensional compact n = (2,2) lattice super yang-mills
  theory with exact supersymmetry},  {\em Phys. Lett.} {\bf B635} (2006)
  218--224, [\href{http://xxx.lanl.gov/abs/hep-lat/0601024}{{\tt
  hep-lat/0601024}}].

\bibitem{Catterall:2004np}
S.~Catterall, {\it A geometrical approach to n = 2 super yang-mills theory on
  the two dimensional lattice},  {\em JHEP} {\bf 11} (2004) 006,
  [\href{http://xxx.lanl.gov/abs/hep-lat/0410052}{{\tt hep-lat/0410052}}].

\bibitem{DAdda:2005zk}
A.~D'Adda, I.~Kanamori, N.~Kawamoto, and K.~Nagata, {\it Exact extended
  supersymmetry on a lattice: Twisted n = 2 super yang-mills in two
  dimensions},  {\em Phys. Lett.} {\bf B633} (2006) 645--652,
  [\href{http://xxx.lanl.gov/abs/hep-lat/0507029}{{\tt hep-lat/0507029}}].

\bibitem{Joseph:2011xy}
A.~Joseph, {\it {Supersymmetric Yang-Mills theories with exact supersymmetry on
  the lattice}},  {\em Int.J.Mod.Phys.} {\bf A26} (2011) 5057--5132,
  [\href{http://xxx.lanl.gov/abs/1110.5983}{{\tt 1110.5983}}].

\bibitem{Elitzur:1982vh}
  S.~Elitzur, E.~Rabinovici and A.~Schwimmer,
  {\it {Supersymmetric Models On The Lattice}},
{\em Phys. Lett.} {\bf B119} (1982) 165.

\bibitem{Kanamori:2012et}
I.~Kanamori, {\it {Lattice formulation of two-dimensional N=(2,2) super
  Yang-Mills with SU(N) gauge group}},
  \href{http://xxx.lanl.gov/abs/1202.2101}{{\tt 1202.2101}}.

\bibitem{Suzuki:2005dx}
H.~Suzuki and Y.~Taniguchi, {\it Two-dimensional n = (2,2) super yang-mills
  theory on the lattice via dimensional reduction},  {\em JHEP} {\bf 10} (2005)
  082, [\href{http://xxx.lanl.gov/abs/hep-lat/0507019}{{\tt hep-lat/0507019}}].

\bibitem{Kanamori:2008bk}
I.~Kanamori and H.~Suzuki, {\it {Restoration of supersymmetry on the lattice:
  Two-dimensional N = (2,2) supersymmetric Yang-Mills theory}},  {\em
  Nucl.Phys.} {\bf B811} (2009) 420--437,
  [\href{http://xxx.lanl.gov/abs/0809.2856}{{\tt 0809.2856}}].

\bibitem{Hanada:2009hq}
M.~Hanada and I.~Kanamori, {\it {Lattice study of two-dimensional N=(2,2) super
  Yang-Mills at large-N}},  {\em Phys. Rev.} {\bf D80} (2009) 065014,
  [\href{http://xxx.lanl.gov/abs/0907.4966}{{\tt 0907.4966}}].

\bibitem{Hanada:2010qg}
M.~Hanada and I.~Kanamori, {\it {Absence of sign problem in two-dimensional N =
  (2,2) super Yang-Mills on lattice}},  {\em JHEP} {\bf 1101} (2011) 058,
  [\href{http://xxx.lanl.gov/abs/1010.2948}{{\tt 1010.2948}}].

\bibitem{Catterall:2010fx}
S.~Catterall, A.~Joseph, and T.~Wiseman, {\it {Thermal phases of D1-branes on a
  circle from lattice super Yang-Mills}},  {\em JHEP} {\bf 1012} (2010) 022,
  [\href{http://xxx.lanl.gov/abs/1008.4964}{{\tt 1008.4964}}].

\bibitem{Ishii:2008ib}
T.~Ishii, G.~Ishiki, S.~Shimasaki, and A.~Tsuchiya, {\it {N=4 Super Yang-Mills
  from the Plane Wave Matrix Model}},  {\em Phys.Rev.} {\bf D78} (2008) 106001,
  [\href{http://xxx.lanl.gov/abs/0807.2352}{{\tt 0807.2352}}].

\bibitem{Ishiki:2011ct}
G.~Ishiki, S.~Shimasaki, and A.~Tsuchiya, {\it {Perturbative tests for a
  large-N reduced model of super Yang-Mills theory}},  {\em JHEP} {\bf 1111}
  (2011) 036, [\href{http://xxx.lanl.gov/abs/1106.5590}{{\tt 1106.5590}}].

\bibitem{Eguchi:1982nm}
T.~Eguchi and H.~Kawai, {\it Reduction of dynamical degrees of freedom in the
  large n gauge theory},  {\em Phys. Rev. Lett.} {\bf 48} (1982) 1063.

\bibitem{Catterall:2011pd}
S.~Catterall, E.~Dzienkowski, J.~Giedt, A.~Joseph, and R.~Wells, {\it
  {Perturbative renormalization of lattice N=4 super Yang-Mills theory}},  {\em
  JHEP} {\bf 1104} (2011) 074, [\href{http://xxx.lanl.gov/abs/1102.1725}{{\tt
  1102.1725}}].

\bibitem{Hanada:2010kt}
M.~Hanada, S.~Matsuura, and F.~Sugino, {\it {Two-dimensional lattice for
  four-dimensional N=4 supersymmetric Yang-Mills}},  {\em Prog.Theor.Phys.}
  {\bf 126} (2012) 597--611, [\href{http://xxx.lanl.gov/abs/1004.5513}{{\tt
  1004.5513}}].

\bibitem{Hanada:2010gs}
M.~Hanada, {\it {A proposal of a fine tuning free formulation of 4d N = 4 super
  Yang-Mills}},  {\em JHEP} {\bf 1011} (2010) 112,
  [\href{http://xxx.lanl.gov/abs/1009.0901}{{\tt 1009.0901}}].

\bibitem{Myers:1999ps}
R.~C. Myers, {\it Dielectric-branes},  {\em JHEP} {\bf 12} (1999) 022,
  [\href{http://xxx.lanl.gov/abs/hep-th/9910053}{{\tt hep-th/9910053}}].

\bibitem{Maldacena:2002rb}
J.~M. Maldacena, M.~M. Sheikh-Jabbari, and M.~Van~Raamsdonk, {\it {Transverse
  fivebranes in matrix theory}},  {\em JHEP} {\bf 01} (2003) 038,
  [\href{http://xxx.lanl.gov/abs/hep-th/0211139}{{\tt hep-th/0211139}}].

\bibitem{Unsal:2005us}
M.~Unsal, {\it Supersymmetric deformations of type iib matrix model as matrix
  regularization of n = 4 sym},  {\em JHEP} {\bf 04} (2006) 002,
  [\href{http://xxx.lanl.gov/abs/hep-th/0510004}{{\tt hep-th/0510004}}].

\bibitem{Ydri:2007ua}
B.~Ydri, {\it {A Proposal for a Non-Perturbative Regularization of N=2 SUSY 4D
  Gauge Theory}},  {\em Mod.Phys.Lett.} {\bf A22} (2007) 2565--2572,
  [\href{http://xxx.lanl.gov/abs/0708.3066}{{\tt 0708.3066}}].

\bibitem{Dijkgraaf:1996tz}
R.~Dijkgraaf and G.~W. Moore, {\it Balanced topological field theories},  {\em
  Commun. Math. Phys.} {\bf 185} (1997) 411--440,
  [\href{http://xxx.lanl.gov/abs/hep-th/9608169}{{\tt hep-th/9608169}}].

\bibitem{Kaplan:2002wv}
D.~B. Kaplan, E.~Katz, and M.~Unsal, {\it Supersymmetry on a spatial lattice},
  {\em JHEP} {\bf 05} (2003) 037,
  [\href{http://xxx.lanl.gov/abs/hep-lat/0206019}{{\tt hep-lat/0206019}}].

\bibitem{Hanada:2010jr}
M.~Hanada, S.~Matsuura, J.~Nishimura, and D.~Robles-Llana, {\it
  {Nonperturbative studies of supersymmetric matrix quantum mechanics with 4
  and 8 supercharges at finite temperature}},  {\em JHEP} {\bf 1102} (2011)
  060, [\href{http://xxx.lanl.gov/abs/1012.2913}{{\tt 1012.2913}}].

\bibitem{Kim:2006wg}
N.~Kim and J.-H. Park, {\it {Massive super Yang-Mills quantum mechanics:
  Classification and the relation to supermembrane}},  {\em Nucl.Phys.} {\bf
  B759} (2006) 249--282, [\href{http://xxx.lanl.gov/abs/hep-th/0607005}{{\tt
  hep-th/0607005}}].

\bibitem{Damgaard:2007xi}
P.~H. Damgaard and S.~Matsuura, {\it {Relations among Supersymmetric Lattice
  Gauge Theories via Orbifolding}},  {\em JHEP} {\bf 0708} (2007) 087,
  [\href{http://xxx.lanl.gov/abs/0706.3007}{{\tt 0706.3007}}].

\bibitem{Takimi:2007nn}
T.~Takimi, {\it {Relationship between various supersymmetric lattice models}},
  {\em JHEP} {\bf 0707} (2007) 010,
  [\href{http://xxx.lanl.gov/abs/0705.3831}{{\tt 0705.3831}}].

\bibitem{Minwalla:1999px}
S.~Minwalla, M.~Van~Raamsdonk, and N.~Seiberg, {\it {Noncommutative
  perturbative dynamics}},  {\em JHEP} {\bf 02} (2000) 020,
  [\href{http://xxx.lanl.gov/abs/hep-th/9912072}{{\tt hep-th/9912072}}].

\bibitem{Armoni:2001br}
A.~Armoni, R.~Minasian, and S.~Theisen, {\it {On non-commutative N = 2 super
  Yang-Mills}},  {\em Phys. Lett.} {\bf B513} (2001) 406--412,
  [\href{http://xxx.lanl.gov/abs/hep-th/0102007}{{\tt hep-th/0102007}}].

\bibitem{Berenstein:2002jq}
D.~E. Berenstein, J.~M. Maldacena, and H.~S. Nastase, {\it {Strings in flat
  space and pp waves from N=4 superYang-Mills}},  {\em JHEP} {\bf 0204} (2002)
  013, [\href{http://xxx.lanl.gov/abs/hep-th/0202021}{{\tt hep-th/0202021}}].

\bibitem{Sugiyama:2002rs}
K.~Sugiyama and K.~Yoshida, {\it {Supermembrane on the PP wave background}},
  {\em Nucl.Phys.} {\bf B644} (2002) 113--127,
  [\href{http://xxx.lanl.gov/abs/hep-th/0206070}{{\tt hep-th/0206070}}].

\bibitem{Das:2003yq}
S.~R. Das, J.~Michelson, and A.~D. Shapere, {\it {Fuzzy spheres in pp wave
  matrix string theory}},  {\em Phys.Rev.} {\bf D70} (2004) 026004,
  [\href{http://xxx.lanl.gov/abs/hep-th/0306270}{{\tt hep-th/0306270}}].

\bibitem{Kato:2011yh}
J.~Kato, Y.~Kondo, and A.~Miyake, {\it {Mass deformation of twisted super
  Yang-Mills theory with fuzzy sphere solution}},  {\em JHEP} {\bf 1109} (2011)
  019, [\href{http://xxx.lanl.gov/abs/1104.1252}{{\tt 1104.1252}}].

\bibitem{Catterall:2000rv}
S.~Catterall and E.~Gregory, {\it {A Lattice path integral for supersymmetric
  quantum mechanics}},  {\em Phys.Lett.} {\bf B487} (2000) 349--356,
  [\href{http://xxx.lanl.gov/abs/hep-lat/0006013}{{\tt hep-lat/0006013}}].

\bibitem{Iso:2001mg}
S.~Iso, Y.~Kimura, K.~Tanaka, and K.~Wakatsuki, {\it {Noncommutative gauge
  theory on fuzzy sphere from matrix model}},  {\em Nucl.Phys.} {\bf B604}
  (2001) 121--147, [\href{http://xxx.lanl.gov/abs/hep-th/0101102}{{\tt
  hep-th/0101102}}].

\bibitem{Giedt:2004vb}
J.~Giedt, R.~Koniuk, E.~Poppitz, and T.~Yavin, {\it {Less naive about
  supersymmetric lattice quantum mechanics}},  {\em JHEP} {\bf 0412} (2004)
  033, [\href{http://xxx.lanl.gov/abs/hep-lat/0410041}{{\tt hep-lat/0410041}}].

\bibitem{Nekrasov:2003af}
N.~A. Nekrasov, {\it {Seiberg-Witten prepotential from instanton counting}},
  \href{http://xxx.lanl.gov/abs/hep-th/0306211}{{\tt hep-th/0306211}}.

\bibitem{Nekrasov:2003rj}
N.~Nekrasov and A.~Okounkov, {\it {Seiberg-Witten theory and random
  partitions}},  \href{http://xxx.lanl.gov/abs/hep-th/0306238}{{\tt
  hep-th/0306238}}.

\bibitem{Ohta:2006qz}
K.~Ohta and T.~Takimi, {\it {Lattice formulation of two dimensional topological
  field theory}},  {\em Prog.Theor.Phys.} {\bf 117} (2007) 317--345,
  [\href{http://xxx.lanl.gov/abs/hep-lat/0611011}{{\tt hep-lat/0611011}}].

\bibitem{Nekrasov:2002qd}
N.~A. Nekrasov, {\it Seiberg-witten prepotential from instanton counting},
  {\em Adv. Theor. Math. Phys.} {\bf 7} (2004) 831--864,
  [\href{http://xxx.lanl.gov/abs/hep-th/0206161}{{\tt hep-th/0206161}}].

\end{thebibliography}
\providecommand{\href}[2]{#2}\begingroup\raggedright\endgroup

\end{document}